# Pre-Flight Calibration of PIXL for X-ray Fluorescence Elemental Quantification

Christopher M. Heirwegh[1†], William Timothy Elam[2], Yang Liu[1], Anusheela Das[1], Christopher Hummel[1], Bret Naylor[1], Lawrence A. Wade[1], Abigail C. Allwood[1], Joel A. Hurowitz[3], Les G. Armstrong[1], Naomi Bacop[1], Lauren P. O'Neil[2], Kimberly P. Sinclair[2], Michael E. Sondheim[1], Robert W. Denise[1], Peter R. Lawson[1], Rogelio Rosas[1], Jonathan H. Kawamura[1], Mitchell H. Au[1], Amarit Kitiyakara[1], Marc C. Foote[1], Raul A. Romero[1], Mark S. Anderson[1], George R. Rossman[4], Benton C. Clark III[5]

1) Jet Propulsion Laboratory, California Institute of Technology, Pasadena, CA 91109 USA
2) Applied Physics Laboratory, University of Washington, Seattle, WA 98105 USA
3) Department of Geosciences, Stony Brook University, Stony Brook, NY 11974 USA
4) Division of Geological and Planetary Sciences, California Institute of Technology, Pasadena, CA 91125 USA
5) Space Sciences Institute, Boulder, CO 80301 USA
† - corresponding author: christopher.m.heirwegh@jpl.nasa.gov, ORCID: 0009-0002-8234-7288

## Table of Contents

# Abstract

The Planetary Instrument for X-ray Lithochemistry (PIXL) is a rasterable focused-beam X-ray fluorescence (XRF) spectrometer mounted on the arm of National Aeronautics and Space Administration's (NASA) Mars 2020 Perseverance rover. To ensure that PIXL would be capable of performing accurate in-flight compositional analysis of martian targets, in situ, an elemental calibration was performed pre-flight on the PIXL flight instrument in a simulated martian environment. The details of this calibration, and implications for measuring unknown materials on Mars are the subjects of this paper. The major goals of this calibration were both to align the spectrometer to perform accurate elemental analysis and, to derive a matrix of uncertainties that are applied to XRF measurements of all elements in unknown materials. A small set of pure element and pure compound targets and geologically relevant reference materials were measured with the flight hardware in a simulated martian environment. Elemental calibration and quantifications were carried out using PIXL's XRF quantification software (PIQUANT). Uncertainties generated were implemented into the PIQUANT software version employed by the PIXL's data visualization software (PIXLISE). We outline in this work, a list of factors that impact micro-XRF accuracy, the methodology and steps involved in the calibration, details on the fabrication of the uncertainty matrix, instructions on the use and interpretations of the uncertainties applied to unknowns and an assessment on the limitations and areas open to future improvement as part of subsequent calibration efforts.

# Introduction

## PIXL and Calibration for Elemental Quantification

The Planetary Instrument for X-ray Lithochemistry (PIXL) is an X-ray fluorescence (XRF) spectrometer mounted on the arm of the National Aeronautics and Space Administration's (NASA) Mars 2020 Perseverance rover (Allwood et al., 2020; Allwood et al., 2021). PIXL delivers a sub-millimeter focused, raster scannable X-ray beam, capable of determining the fine-scale distribution of elements in martian rock and regolith targets. PIXL was conceived following the work by Allwood et al. (2009) that demonstrated how micro-XRF elemental mapping could reveal the fine-textured chemistry of layered rock structures of ~3,450-million-year-old Archean stromatolitic fossils. Their work not only pushed back the accepted earliest possible window for the beginning of life on Earth, but also demonstrated that significant science return might be possible through XRF mapping. PIXL was proposed, selected, and developed to carry out petrologic exploration that provide the paleoenvironmental context required in the search for biosignatures on Mars, analogous to Allwood et al.'s earlier work.

PIXL was developed to analyze inorganic elemental rock signatures for elements of atomic mass of Na and higher on the periodic table. The calibration of PIXL to perform accurate quantitative analysis of major, minor and trace levels of these element constituents, as found in rock and regolith materials on Mars, is the focus of this work.

Calibration is defined for this work as having two functions. One is to align the elemental response of PIXL's analytical output to give the best possible reproduction of the true elemental abundance in a sample, across the broadest range of unknown materials. The second is to characterize constraints on accuracy so that quantified abundances may be reported with accurate estimates of uncertainties.

For the past 50 years, NASA has incorporated XRF spectrometer instrumentation in the payloads of numerous lander and rover spacecrafts sent to Mars. Each instrument has contributed both to the exploration of Mars' geological history (e.g., McSween et al., 2009; McLennan et al., 2019) and to the search for evidence of habitable conditions with the potential to support an ancient martian biosphere (e.g., Grotzinger et al., 2005; Knoll et a., 2005; Ruff et al., 2011). Each instrument was calibrated to perform elemental analysis. Spectrometers, and their calibrations, sent to Mars include the Viking lander X-ray Fluorescence Spectrometer (XRFS) (Clark et al., 1977), the Pathfinder Sojourner Rover's Alpha-Proton X-ray Spectrometer (ApXS) (Rieder et al. 1997b, 1997a), the Mars Exploration Rovers, Spirit and Opportunity's, Alpha-Particle X-ray Spectrometers (APXS) (Gellert et al. 2006; Rieder et al. 2003), and the Mars Science Laboratory (MSL) Rover, Curiosity's APXS (Campbell et al. 2012). Calibrations in all cases relied upon measurements of geological reference materials (GRMs) to assess the spectrometer's elemental response and to derive limits on accuracy.

Like its predecessors, PIXL supports the exploration of Mars' geology and past habitability (Farley et al., 2020). PIXL's mapping capabilities are further being utilized to support the search for targets carrying high potential for preservation of biosignatures. The fast turnaround on producing scientific data products offered by PIXL further informs the Mars 2020 project in their initiative to cache samples of cored rock and loose regolith. The return of these samples to Earth by NASA and the European Space Agency's planned Mars Sample Return (MSR) campaign would permit more rigorous scientific study in support of uncovering the mysteries of Mars' past (NRC, 2011; NASEM, 2023).

Several key subsystems in PIXL (see Figs. 2 & 5 in Allwood et al., 2020) operate in concert during a measurement to provide PIXL's downlinked X-ray and multispectral image data products. These include the X-ray subsystem (Figs. 16, 17 & 20 in Allwood et al., 2020), the optical fiducial sub-system (OFS) (Fig. 1 in Klevang et al., 2023, Figs. 6, 24 &25 in Allwood et al., 2020), and the mechanically driven hexapod positioning system (Fig. 8 in Allwood et al., 2020). The X-ray subsystem, comprised of the X-ray source and detector components (Fig. 6.b) in Allwood et al, 2020), enables elemental analysis of the chemical composition of the interrogated rock. The X-ray source includes a Moxtek™ custom-designed thick Rh anode tube system mated to a XOS® glass polycapillary X-ray focusing optic. When run under nominal conditions of 20 μA and 28 kV, this source delivers a sub-mm focused polychromatic X-ray beam to the rock surface. Two Ketek GmbH Vitus H50 detectors, mounted on either side of the optic in near-backscatter geometry, record the raw X-ray signal emitted by the rock. The detectors are each tilted 20° to point directly at the X-ray focal spot and, as viewed from the focal location, are angled 20° down from the boresight of the beam. Under these conditions, common rock-forming major, minor and trace elements are detectable by their K X-ray line emissions (11 ≤ Z ≤ 42); higher Z elements may be detected by their less intense L X-ray emissions. Digital signal processing electronics (Figs 3 & 9 in Allwood et al., 2020) on the rover body convert the pre-amplifier signals transmitted from the PIXL sensor assembly through a flex cable on the arm, into counts per X-ray energy in a histogram-like spectrum. Raw PIXL spectra, returned to Earth are then turned over to in-house-developed spectra processing software for analysis. The PIXLISE graphics user interface software provides a workspace to correlate target elemental maps with visible camera images and derive trends in target mineralogy and composition (Allwood et al., 2020; Barber and Davidoff, 2022; Liu et al., 2022; Tice et al., 2023). PIXLISE uses PIXL's quantitative XRF software routine (PIQUANT) to fully quantify individual pixels of elemental map data (Heirwegh et al., 2022).

## PIXL's XRF Quantification Software

PIQUANT is an open source (Elam et al., 2022) C++ X-ray quantification code that utilizes a physics fundamental parameters (FP) approach to converting raw X-ray spectral elemental peak areas into quantified elemental abundance. The FP approach considers all physical parameters that affect the emission of X-rays including: measurement geometry, instrument construction materials, X-ray transmission through the optic and detector windows, environmental conditions and database values for the interaction of X-rays within materials (Heirwegh et al., 2022). PIQUANT calculates the line shape for each elemental peak as a Lorentzian using the natural line width, then adds in calculated incomplete charge collection artifacts (Scholze and Procop, 2009). The peak plus artifacts are then broadened by a Gaussian distribution using the energy-dependent detector resolution. The X-ray bremsstrahlung continuum is treated using a combination of models. Background is calculated using, 1) physics first principles for the 0.5 – 7 keV region, SNIP fit removal (Ryan et al., 1988) for the 7 – 28 keV region and an escape "shelf" feature at 0.5 – 1.5 keV representing Compton scatter escaping from the detector (Michel-Hart and Elam, 2017). PIQUANT calculates the full XRF spectrum, peaks and background, using the above information and attempts to match the calculated spectrum to the full measured spectrum from each PIXL measurement spot.

Development of XRF quantifications codes, in general, can utilize one of several possible approaches, which include the FP approach, the Monte Carlo approach or the reference-based approach. The decision to code PIQUANT to rely on the FP approach was made during instrument proposal development (2012-2013) for several reasons. First, the Monte Carlo approach, though a strong candidate, was considered to require too much computational power and require too long a processing time to provide chemical abundance feedback on an operational tactical timeline. For example, initial estimates of bulk chemical abundances, called "Quicklooks", are required by the Mars 2020 project in less than 30 minutes after receipt of data products from Mars. The reference-based approach, though faster, relies upon calibration, in the first step as defined earlier, by having the spectrometer measure hundreds of reference materials. For such an approach, data from many measurements are used to plot curves of the spectral peak area response, as a function of the known elemental concentration of the GRM, for each element. Subsequent measurements of unknown materials are processed by interpolating from these plots, for each element, using the fitted peak areas as input. While such a method can yield generally reliable results, the PIXL team considered that there are insufficient suitable Mars-analogue materials to build such a large reference material base. Additionally, this method would have eaten up X-ray tube life, a finite resource needed for measuring targets on Mars, and was therefore less favored to the FP approach. The applied FP approach was therefore deemed the appropriate choice. Developing the software in-house also permitted complete control over the selection of FP databases incorporated in the software.

With PIQUANT's FP approach, compositions of unknown materials are derived iteratively through comparing and matching calculated spectral lineshapes to raw spectral data and adjusting the elemental composition until matching is realized. Lineshapes are calculated using a starting "guess" at the composition that assumes all user identified elements are equal in abundance. This guess is inherently incorrect for natural materials and the calculated lineshapes do not match the raw data, so the program iterates to improve on the guess at composition. With each iteration, new estimates of elemental concentrations are produced, adjusted in a direction that steadily shrinks the difference between the calculated vs measured lineshapes. A best estimate representation of the elemental abundances of the unknown is achieved once the calculated vs measured lineshapes agree to within 0.1% difference between the second last and last iterations, for all elements. Through this calculation the software assumes that the matrix mixture of elements is purely homogeneous, a condition that must be imposed upon single spot measurements. Only those elements that are visible to PIXL ($Z \geq 11$) are varied in the fit and visible elements are stoichiometrically linked to oxygen and/or carbonate constituents as defined by the user. Further details on this iteration process and software architecture are found elsewhere (Heirwegh et al., 2022).

## Factors that Affect PIXL Micro-XRF Measurement Uncertainties

One of the key outcomes to this work is to have knowledge of the final uncertainties that must be applied to each and every element quantification produced by PIXL and the PIQUANT software. Each elemental uncertainty comprises numerous factors that can be grouped as affecting either measurement precision or measurement accuracy. The uncertainties that are derived in this work for PIXL combine both groups. Our approach bears some resemblance to the approaches used to calibrate the MER (Gellert et al, 2006) and MSL (Campbell et al., 2012) APXS flight equivalent units (FEU's); these calibrations were then extended to the flight hardware. Both of the APXS calibrations relied on assessing accuracy using the spread in results obtained from measuring GRMs. In particular, Gellert et al. (2006) reported each elemental uncertainty using the mean deviation of APXS quantifications relative to reference values assessed across multiple GRMs. Campbell et al., (2012) performed a similar assessment with uncertainties reported at a 2-sigma range. Due to the similarity of the spectrometers, the Gellert et al. (2006) MER APXS uncertainties are considered approximately accurate in their application to the MSL APXS data (Gellert et al., 2015). Though these uncertainties reflect a realistic spread of measurements that might be obtained from measurements of materials on Mars, most APXS results are reported with only the peak-area statistical, or precision, uncertainties given. The uncertainties quoted in this work may thus appear inflated when compared to APXS statistical uncertainties but are remarkably similar to the APXS accuracy-based uncertainties reported by Gellert et al. (2006).

Factors affecting precision of a measurement include Poisson statistical variations in the emission of X-rays from atoms, variations in X-ray tube emissions (i.e., tube current and voltage), and instrument standoff reproducibility using the optical fiducial subsystem (OFS). Details on the pre-flight calibration of the OFS subsystem has been discussed elsewhere (Klevang et al., 2023).

Measurement accuracy is affected by environmentally induced temperature swings; instrument tip/tilt used to point the instrument when scanning the X-ray beam over a targeted rock surface; X-ray yield losses due to fine surface roughness and porosity; topography variation effects on standoff distance; differences in viewing geometry between detectors that is induced by topography and fine scale material heterogeneities, which includes cases where the X-ray beam excites, or straddles, several material phases at once. Although temperature swing affects are largely mitigated through on-board heating mechanisms, extreme swings might still affect X-ray source output intensity and cause variations in the energy-channel calibration of recorded X-ray spectra. In addition, the extreme thermal environment also affects the stability of the platform which PIXL is mounted to, turret and robot arm of Perseverance. Throughout a PIXL scan, the thermal environment changes up to 80C, causing the platform to positionally drift. These effects are autonomously accounted for by PIXL using terrain relative tracking technology, keeping the X-ray on point to the targeted region (Klevang et al 2024).

We note that tilt uncertainty and topography variation effects on the OFS in maintaining constant standoff distance were not factors in the calibration reported here because calibration target materials were translated in front of a fixed PIXL sensor assembly and aligned with a boresight that was always perpendicular to the target surfaces.

Uncertainties introduced by PIQUANT may also affect accuracy. Software sources include empirical algorithms used to model X-ray emission from the Rh tube anode, limitations in peak lineshapes and background subtraction algorithms applied to the raw data and, any discrepancies in the FP databases used by PIQUANT to convert X-ray peak intensity to element abundance. Descriptions of the FP approach and databases used by PIQUANT are described elsewhere (Heirwegh et al., 2022). Examples of discrepancies observed in FP databases, including those used by PIQUANT, have been outlined by the International Initiative on X-ray Fundamental Parameters (IIFP) (Campbell et al., 2017). Many of these relate to limitations in theoretical models or data used to derive FP databases.

### Steps in the Elemental Calibration and Role of the PIQUANT Software
Several steps are involved in the calibration. The first is to derive the relative transmission of photons through the optic as a function of energy (or wavelength) to provide the optic response (OR) relationship. This relationship is used by PIQUANT as part of the FP approach making elemental quantification possible. The most straightforward way to derive the OR would be a line-of-sight measurement of the tube output with an X-ray detector both with and without the optic in place. A ratio comparison of the two detector-recorded beam outputs would then yield an estimate of the OR as a function of X-ray energy. In practice, this approach is complicated by the overwhelming flux emitted by the focused and unfocused sources when measured in full by a semiconductor detector. As well, taking the channel-by-channel ratio of the two raw responses is plagued with large divergences about characteristic lines, any diffraction peaks present, counts at low energies when window absorption has effect, and the drop of spectral counts to zero intensity at both low and high energy boundaries.

As summarized elsewhere (Heirwegh et al., 2022), others have utilized several methods to derive the OR, all of which identified the challenges associated with their methods. Due to the complex construction and geometry of the X-ray optic, it is not possible to determine the OR analytically with sufficient accuracy. Monte Carlo modelling of the setup is also difficult and would require more work to validate this approach. We have utilized an indirect means of measuring and processing the OR, which is described below, in brief, in the Methods section. A more extensive description of the OR derivation is being prepared as part of a future publication. Once the OR is calculated and incorporated into the analytical software algorithm, the FP approach is then able to perform elemental quantification.

Despite best efforts, current methods used to derive the OR, including our own, may be subject to significant errors in certain energy ranges. As well, the FP approach may introduce significant global errors in the quantification of certain elements. This necessitates the use of additional coefficients, defined as element calibration factors (ECFs), one assigned to each element, to improve the accuracy in analysis. ECFs, derived as the second step in the calibration, are a set of fixed coefficients that are utilized for every quantification task. The functionality of the ECF is similar to the adjustment constants used in other X-ray fluorescence analytical applications such as H values for particle induced X-ray emission (PIXE) analysis and K ratios in electron-induced energy dispersive X-ray spectroscopy (EDS).

Generally, ECF corrections applied to quantification results do not impart more than a 30% correction and most are on the order of 10%. Very approximately, the improvement imparted to analysis by inclusion of ECFs is to shrink the potential 10 – 20% error in quantifying major and minor element constituents found in geological materials down to 3 – 10%. Without ECFs, in theory, if the FP quantification of an element is perfectly accurate, such as in the case that all FPs and calculation steps are fully known, then an ECF derived for that element will be unity (i.e. = 1.0). The multiplication of this number by the FP equation would, in this case, impart no change to the final quantified result. By extension, any global discrepancy in the FP approach for a given element will produce an ECF that has an equivalently-sized departure from unity. This condition is necessary as it permits correction for the discrepancy when quantifying elements in unknown materials.

With both the OR and ECFs in place, the PIXL instrument-plus-PIQUANT software as a complete analytical system is effectively calibrated and ready to perform analysis of unknown materials. We consider a third and final major step to the calibration process, which is an assessment of the accuracy or uncertainties associated with the calibration. This step is accomplished by measuring GRMs, treating them as unknown materials, and assessing the PIQUANT calculated quantification results against reference or certificate values. This final calibration step is used to assess the alignment of the calibration and generate uncertainties as a function of both atomic number and element concentration in the sample. This provides the framework used to build an uncertainty mesh that is used by PIQUANT to produce uncertainties for unknown materials.

For our in-lab calibration, additional uncertainties (affecting accuracy) are introduced through use of GRM certificate or reference compositions, which may not accurately reflect the composition of the measured GRM aliquot. For example, effects from uncertainties originating from techniques used by other labs are a significant portion of a GRM's uncertainty. Also, micro-level heterogeneities in powder GRMs and micro-levels of surface roughness in the press-powder materials will inflate the possible errors of quantifying these materials.

PIQUANT FP software is used in each of the three calibration steps and to process PIXL FM measurements performed on a limited standard set. The OR is derived using measurements of a PTFE (Teflon™) "blank" material. ECFs are derived using 14 pure element and pure compound materials, either in solid or pressed powder form. The same pure materials plus an additional set of glass, powder and mineral GRMs are used as part of the third calibration step. Details on the derivation of each of these three steps is outlined in the Methods section.

## Overview of the PIQUANT Processing Algorithm

We now describe some of the formalism that is germane to the processing sequence used by the PIQUANT software. Much of this discussion builds on the description of the software and FP equations outlined in Heirwegh et al. (2022).

By definition, when using the physics FP approach, if we assume that we know the true composition of a sample and there are no uncertainties associated with this approach, then PIQUANT's calculation of expected peak area $(P'_{i,calc})$, associated with element $(i)$, will match exactly with the raw data peak area for that element $(P_{i,data})$, as measured by PIXL. This is shown in equation 1.

$$P_{i,data} \stackrel{\text{def}}{=} P'_{i,calc}(P, S, T, ECF) \qquad (1)$$

$P'_{i,calc}$ is an FP equation that includes contributions from both primary $(P)$ and secondary $(S)$ X-ray fluorescence, as given in equations 1 and 2 of Heirwegh et al., (2022). Equations $P$ and $S$ in turn comprise: all X-ray fundamental parameters, instrument geometric parameters and elemental concentrations. $P'_{i,calc}$ also includes correction for X-ray transmission $(T)$ through the optic, as taken from the OR as a function of X-ray wavelength $(\lambda)$, or energy. $ECF$s are also included in this expression. $P'_{i,calc}$ is therefore a function of each of the four aforementioned major variables $P, S, T, ECF$.

$P'_{i,calc}(P, S, T, ECF)$, the right side of eqn. 1, is expanded further (eqn. 2) to illustrate how these variables are convolved together.

$$P_{i,data} = (P_{i,calc} + S_{ij,calc}) \cdot \left\{ \int_{\lambda_0}^{\lambda_{abs,i}} T(\lambda) d\lambda \right\} \cdot ECF_i \qquad (2)$$

Here, $P_{i,calc}$ is the primary fluorescence intensity for element $i$, $S_{ij,calc}$ is the calculated secondary fluorescence emitted by element $i$ as a result of excitation by X-ray emissions from element $(j)$. Contributions from optic transmission as a function of X-ray wavelength are integrated across the full wavelength range responsible for exciting element $i$. This extends from the highest energy limit, or shortest wavelength $(\lambda_0)$, to the absorption edge wavelength for that element $(\lambda_{abs,i})$. ECFs are also specific to each element $i$. We again note that concentration $C_i$ for element $i$ is found inside the expressions for primary and secondary yield. PIQUANT works by minimizing the difference between the left and right sides of this equation and does this by adjusting the concentration of each element in a sample, as indicated above.

In the Methods section, we outline the selection of the target materials set assembled for the calibration, the chamber apparatus and conditions used to record PIXL data of the target materials, the steps taken to derive the optic response and ECFs and finally the assessment of the calibration and uncertainties associated with each PIXL measurement. This manuscript provides full details on the elemental calibration that have been presented summarily prior to this work (Heirwegh et al. 2021; Heirwegh, 2023).

## Methods

### Selection of the Elemental Calibration Target Set

The calibration materials gathered were grouped into three sets, as shown in Table 1. The first set comprises the pure element and pure compound targets used to derive the OR and ECFs, steps 1 and 2 of this calibration. These are: high-purity Teflon (PTFE), NaCl, $MgCO_3 \cdot H_2O$, $Al_2O_3$, $SiO_2$, ZnS, KBr, $CaF_2$, Ti, Fe, Ge, Y, Zr, $BaZrO_3$ and CeO. The choice in materials selected for ECF derivation was made to provide enough spacing in atomic number that unmeasured elements could be interpolated or extrapolated. Through use of these targets, 17 element ECFs are realized. These pure targets were also selected to support future assessment of secondary spectral artifacts such as peak tailing.

The second set includes geological materials that represent a spread in differing material types found on Earth. The set includes: fused basaltic glasses (BHVO2-G, BIR-1G, BCR-2G), trace-element doped glass (SRM® 610), lake sediment (LKSD-4), gypsum (Gyp-B), carbonatite (COQ-1), calcium phosphate (SRM 694), dunite (6 NIM-D), marine sediment (JMS-2) and phlogopite (Mica Mg). These materials were selected to reproduce the X-ray response as might be observed from the possible spread of igneous and sedimentary type targets, thought to be present at Jezero Crater on Mars. This second set, the "GRMs", were utilized in the third calibration step.

The third set is the calibration target set now mounted on Perseverance (see figs. 4 and 13 in Allwood et al., 2020). The set consists of: PTFE, BHVO2-G, SRM® 610, donated by Clay Davis (NIST), a piece of natural scapolite mineral, donated to this project from the private collection of G. R. Rossman (co-author on this work), and a fused silica puck substrate lined with 200 um wide Ni and Cr metal strips. These materials are noteworthy for their ongoing use by PIXL to check for the consistency of the existing calibration as the X-ray system ages. This target set may also be utilized to rederive the calibration should the need someday arise. Only the scapolite measurement data from this target was utilized in the pre-flight calibration in the third step. Data were recorded pre-flight for the rover calibration target to provide a link of the ground calibration to all subsequent measurements of the calibration target made on Mars.

Reference materials were chosen for the rover target to provide a broad spread of elemental representation. BHVO2-G provides most major and minor elements of geological interests: Na, Mg, Al, Si, K, Ca, Ti, Cr, Mn, Fe, Ni, Cu, Zn Sr and Zr. Scapolite provides many of these elements as well but most notably, S and Cl, neither of which are found in BHVO2-G. The SRM® 610 sample also provides several major element constituents but most notably, abundant higher-Z trace elements, including those emitting strong L X-ray lines in addition to K lines. The centrally located cross-hair puck is used as part of beam location and camera-X-ray boresight position co-registration and is discussed in Allwood et al. (2020).

**Table 1:** Target materials set used as part of the pre-flight ground calibration of PIXL. Targets are sorted into groups of "Pure", simple spectra materials, "GRM" for geologically relevant materials, and "Rover", from the on-board calibration target. Information on ECF elements examined, time measured by PIXL, the reference source for elemental composition and the full chemical name of the targets are shown in columns 2 through to 6, respectively.

| ID | ECF element | Supplier | Time (*min*) per det. | Comp. source | Full name |
| --- | --- | --- | --- | --- | --- |
| *Pure* | | | | | |
| PTFE | | IFP | 120 | Stch | Polytetrafluoroethylene [-($C_2F_4$)$_n$-] |
| NaCl | Na, Cl | IO | 5 | Stch | sodium chloride |
| $MgCO_3 \bullet H_2O$ | Mg | FA | 5 | Stch | magnesium carbonate |
| $Al_2O_3$ | Al | IO | 5 | Stch | alumina window |
| $SiO_2$ | Si | EO | 5 | Stch | silica window |
| ZnS | S, Zn | IO | 5 | Stch | zinc sulfide window |
| KBr | K, Br | IO | 5 | Stch | potassium bromide window |
| $CaF_2$ | Ca | IO | 5 | Stch | calcium fluoride window |
| Ti | Ti | AE | 5 | Stch | titanium |
| Fe | Fe | AE | 5 | Stch | iron |
| Ge | Ge | AE | 5 | Stch | germanium |
| Y | Y | AE | 5 | Stch | yttrium |
| Zr | Zr | AE | 5 | Stch | zirconium |
| $BaZrO_3$ | Ba, Zr | Aldrich | 5 | Stch | barium zirconate |
| CeO | Ce | AE | 5 | Stch | cerium oxide |
| *GRM* | | | | | |
| BHVO2-G | | USGS | 120 | Cert | Basaltic Hawaiian Volcanic Ocean Glass |
| BCR2-G | | USGS | 120 | Cert | Basalt, Columbia River |
| BIR1-G | | USGS | 120 | Cert | Icelandic Basalt Glass |
| SRM® 610 | | NIST | 120 | Jochum | trace elements in glass |
| 6 NIM-D | | SARM | 120 | Cert | dunite |
| Gyp-B | | DOM | 120 | Cert | gypsum |
| COQ-1 | | USGS | 120 | Cert | carbonatite |
| JMS-2 | | GSJ | 120 | Cert | marine sediment |
| SRM 694 | | NIST | 120 | Cert/JPL | Western phosphate rock |
| LKSD-4 | | CANMET | 120 | Cert/JPL | Lake sediment |
| Mica Mg | | SARM | 120 | Cert | Phlogopite |
| *Rover* | | | | | |
| R-scapolite | | Rossman | 120 | EPMA | GRR 2409 Marialite, Morogoro, Tanzania |
| R-PTFE | | IFP | 120 | Stch | Polytetrafluoroethylene |
| R-BHVO2-G | | USGS | 120 | Cert | Basaltic Hawaiian Volcanic Ocean Glass |
| R-SRM® 610 | | NIST | 120 | Cert | trace elements in glass |

**Abbreviations:**
Stch – calculated from stoichiometry, Cert – certificate values
SRM - standard reference material
NIST - National Institute of Standards and Technology, Gaithsburg, MD, USA
SARM - Service d'Analyse des Roches et des Mineraux, CRPG-CNRS, Nancy, France

DOM - Domtar Inc.
EO – Edmund Optics
IFP - Industrial Fluoroplastics, Inc.
IO – ISP Optics

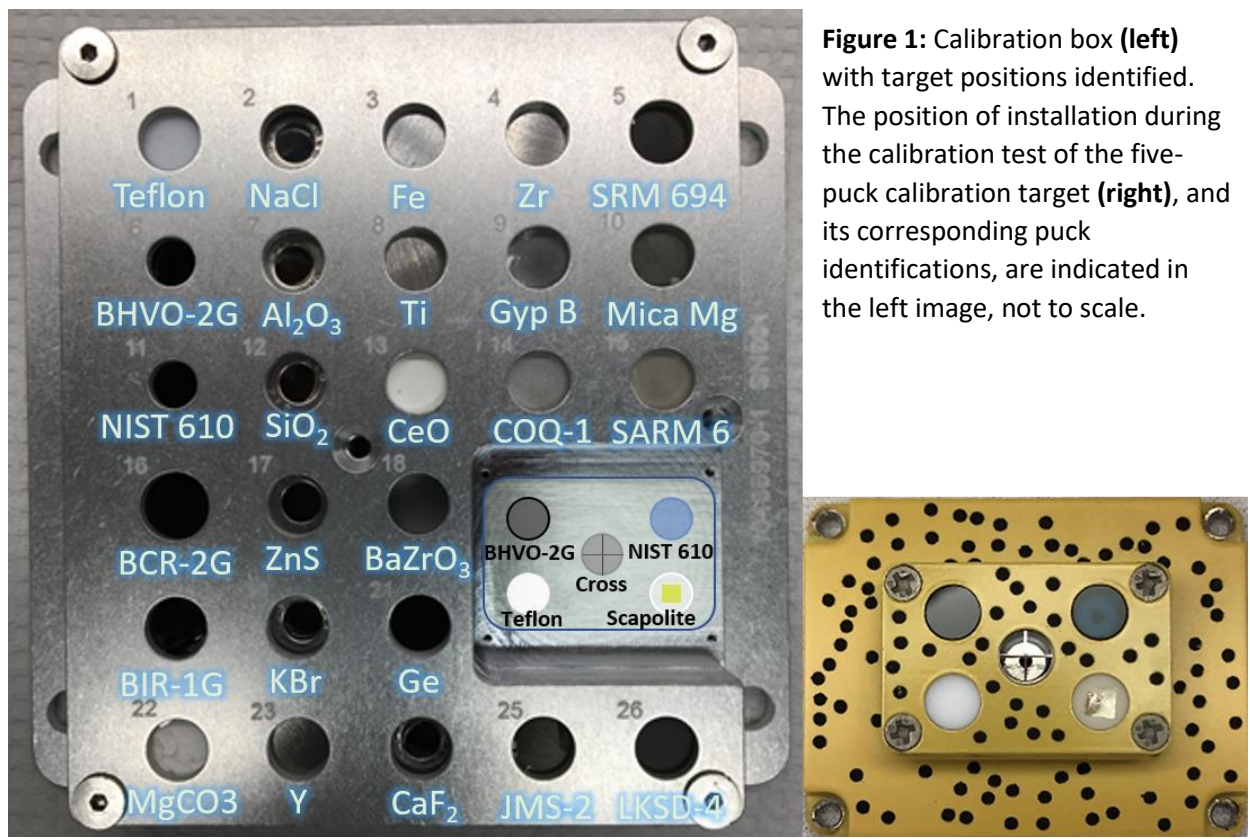

**Figure 1:** Calibration box **(left)** with target positions identified. The position of installation during the calibration test of the five-puck calibration target **(right)**, and its corresponding puck identifications, are indicated in the left image, not to scale.

## Preparation of Target Materials

Several steps of preparation of the samples were taken to ensure that targets were contamination free and presented flat geometry with minimal deviations in density. Additional steps were taken to ensure that reference compositions reported elemental abundances with as much accuracy as possible.

GRMs were heated under vacuum to remove unbound water components. All powder materials were dried in a vacuum oven for 5 days at 140°C to remove unbound water. Water loss was measured by weighing the samples pre- and post-drying. Unbound water accounted for only a few percent weight of the standards except for Gyp-B, which lost ~14% of its weight as water. This measured loss is consistent with the conversion of gypsum (20.9 wt.% $H_2O$) to bassanite (6.2 wt.% $H_2O$). The heating of Gyp-B to 140°C may have produced a nearly anhydrous form of Gypsum (i.e. << ½$H_2O$) but likely, rapid re-hydration to bassanite (½$H_2O$) had occurred (Harrison, 2012) from periodic and short exposures of the calibration materials to ambient air during transfer between preparation stages. As the true hydration state is not fully known and given that this would not be known from a Mars PIXL measurement, we simply assumed the gypsum to be purely anhydrous in this calibration.

The dried powders were then pressed into pellets using a stainless-steel holder and multi-ton pellet press in the ambient environment. To minimize possible water re-absorption during preparation and storage, pellets were subsequently dried at 105°C for several hours and then stored for extended time at 70°C under vacuum. Pressing in this manner also produced targets that possessed flat measurement surfaces.

Some of the USGS glass standards were machined slightly to fit into the slots of the holder shown in Fig. 1. All solid targets came pre-polished from suppliers, thereby ensuring a flat measurement face. Care was taken to ensure that fingerprint contamination of the surfaces was avoided. All targets, including the 5-puck rover calibration target, were mounted with identical recesses of their top faces. This was to ensure that the X-ray sensor head standoff (the 'Z' direction) could be maintained across a full X-Y motion of the sample stage.

### Preparation of Reference Compositions

Several steps were involved in the preparation of the reference compositions used in this work. First, GRM concentration values were tabulated using supplier certificates, [GeoReM](#) (Jochum et al., 2005) recommendations or publications, whichever appeared to be the most complete and current. Sources were summarized in Table 1. Next, targets that experienced a loss of unbound water, due to drying, were renormalized to remove the unbound water and ensure a 100% geochemical sum of the remaining elements. The third step involved a case-by-case assessment for the treatment of elements and compounds that are not visible to PIXL that are found in a number of GRM targets. Fourth, preliminary analysis of PIXL data of these targets revealed that for several targets, entries for the concentrations of element S were missing. This prompted new non-PIXL compositional analysis of these targets, conducted at JPL, to verify the concentrations. The details of all augmentations made to refine compositional reference values are summarized in Table 2 and explained as follows.

The first two steps were straightforward and are not discussed further. For the third step, we consider that PIXL cannot directly detect and quantify "invisible" elements (e.g. B, Be, F) or compounds (e.g., water, hydroxyl groups, organics and carbonates) found in many GRMs. Only oxygen and/or carbonate ($CO_3$) will be quantified through stoichiometric linkage to the "PIXL-visible" elements.

We also considered that FP quantification relies upon accurate knowledge of a target's matrix composition to correct for X-ray absorption affects. Leaving the XRF-invisible components out of our target element description runs the risk of diminishing the accuracy in this correction, specifically if the quantity is of relatively high abundance (≥ 5 wt.%) in the sample. PIQUANT can add invisible constituents other than oxygen or carbonates as fixed quantities that are not iterated as part of the FP calculation. Only invisible elements of abundances higher than 5 wt.% were included in our compositional input files.

Operationally, this approach was justified in that we would likely not be able to discern low abundance invisible constituents with PIXL or any other instrument. However, larger abundance constituent identities (e.g. 20 wt.% $CO_3$) might be inferred with supporting instrument measurements, combined with significantly low geochemical sums derived by PIXL. We note that for analytical consideration that PIXL does not enforce geo-chemical summing to 100 wt.% which, differs from the approach used in APXS analyses.

Further steps were taken to identify elemental constituents suspected as being present in the GRM, despite not being reported anywhere in the reference composition literature. A substantially large loss-on-ignition (LOI) component at 33.5 wt.% was reported for COQ-1. This was interpreted to be pure $CO_3$ (Steve Wilson, USGS, private communication). Computationally, this was enforced by stoichiometrically linking elements, Mg, Ca, Mn, Fe and Sr to $CO_3$ groups using PIQUANT's carbonates feature. For LKSD-4 the 500°C LOI of 40.8 wt.% was broken down into water, 6.55 wt.% removed from drying, and an un-defined mixture of carbon, carbon-oxygen and oxygen compounds having a total sum of 34.3 wt.%. As the certificate indicated 17.7% pure carbon, the remaining amount of the 34.3 wt.% was denoted O at 16.6 wt.%. The intent was to imply a mixture that loosely represents the possible carbonate or carbon-gas present in the material. This latter point was relevant as far as ensuing that X-ray absorption in the target material could be calculated as accurately as possible, which benefits the accuracy in quantifying the XRF-visible components.

**Table 2:** specific treatments applied to certificate values to get final element concentrations and reporting used in this work.

| GRM ID | Notes |
|---|---|
| LKSD-4 | • 500°C LOI (mass loss on ignition) cited as 40.8 wt.%, broken down as: 6.55% $H_2O$ (considered removed by drying), 17.7% pure C (certificate) and 16.6% O (remainder) |
| | • reference pure S content of 0.99 wt.% verified at JPL via gravimetric analysis at 1.09(±0.02) wt.% and assumed to have stoichiometry of $SO_3$ (equiv. 2.47 wt.%). |
| | • Geo-chemical normalization to 100% of reference compositions raised all wt.%s by 5%. |
| JMS-2 | • $H_2O$ content was considered to be completely removed by heating as above. |
| COQ-1 | • Assumed un-defined wt.% component to be pure carbonate ($CO_3$) |
| | • We set carbonates flag in PIQUANT which assigns appropriate $CO_3$ stoichiometry to elements Mg, Ca, Mn, Fe and Sr as giving $MgCO_3$, $CaCO_3$, $MnCO_3$, $FeCO_3$ and $SrCO_3$. |
| Gyp-B | • LOI 22.85% with 17.8% of this being $H_2O$ and 5% $CO_2$. We have removed both $H_2O$ and $CO_2$ from the sample and have renormalized the sample concentrations |
| Mica Mg | • Water and 2.85 wt.% F left out as we would not observe either with PIXL. |
| 6 NIM-D | • Assumed 1.75% LOI in certificate is mostly $H_2O+$ and $H_2O-$, both left out. |
| SRM 694 | • 3.2% F left out as we would not observe it with PIXL. Observed S and measured content via combustion analysis. Pure S determined to be 1.0% |
| $MgCO_3$ | • Considered chemically to be monohydrate: $MgHCO_3 \cdot (OH)$ |

Another target that required reference compositional preparation was NIST SRM 610, an artificial glass doped with about 60 elements of approximate 400 ppm trace abundance, each. The certificate for this material provides a pre-doped composition for the four major element-oxides constituents that comprise the glass ($Na_2O$, $Al_2O_3$, $SiO_2$, CaO); the abundance of these sums to 100%. As well, only a select number of trace dopants were provided in the certificate. A more complete description of the glass dopant concentrations were extracted from Jochum et al., (2011), second column of Table 8. To renormalize all constituents to yield a 100% geochemical sum, the sum of the trace element abundances was subtracted from the total undoped glass abundance. The glass constituent compositions were then renormalized so that the sum of the glass constituents plus all trace elements sum to 100 wt.%.

In two GRMs (NIST SRM 694 and LKSD-4), sulfur K$\alpha$ peaks were found present in PIXL XRF spectra but sulfur was not reported in the certificate analyses. Sulfur content was analyzed at the JPL Analytical Chemistry Laboratory via combustion analysis using a LECO CS844 elemental analyzer. The sulfur recovery was further verified using an anhydrous $CaSO_4$ standard. Loss of volatile elements was investigated using a TA Q500 TGA thermogravimetric analyzer. SRM 694 and LKSD-4 were found to contain 1.00±0.04 wt.% (1$\sigma$, $n$ = 6) and 1.09±0.02 wt.% (1$\sigma$, $n$ = 6), respectively. Uncertainties for these averages are based on six independent analyses of each material. S abundances were integrated with renormalization to become part of the certificate composition. As an exercise to cross-validate our S analytical methods and results with that reported in other composition references, we performed further combustion analyses on Gyp-B and JMS-2. Concentrations of S were found to be: 20.0±1.0 wt.% (±1$\sigma$, $n$ = 7) sample (certificate: 21.3%) and 0.276±0.015 wt.% (±1$\sigma$, $n$ = 7) sample (certificate: 0.29%), respectively. We considered the approximate 5% difference in our results relative to certificate values in these two samples to be small enough to validate our chemical analyses of SRM 694 and LKSD-4.

As carbonate is highly relevant to many sedimentary rocks and altered igneous rocks, we prepared a reference pellet of pure anhydrous $MgCO_3$ powder. Its composition was estimated to be monohydrate reflecting partial rehydration from pure anhydrous state due to periodic exposure to air and assigned the following stoichiometry: $MgCO_3 \cdot H_2O$.

Different fragments of scapolite were polished for the FM calibration-target and for the in-house standard set. These fragments were derived from a gem-quality fragment (internal catalog, GRR 2409). The major- and minor-element chemistry of the in-house scapolite were determined in 1991 at Caltech on a different piece of GRR 2409 using a JEOL 733 electron probe micro-analyzer (EPMA). For these measurements, electron beam settings of 25 kV and 10 nA were used to collect characteristic X-rays of F, Na, Al, Si, S, Cl, K, Ca, Ti, Fe, and Sr from five different spots on the sample. The electron beam was also defocused to a diameter of 5 µm. To calibrate the JEOL 733 EPMA, natural and synthetic standards were used, including fluoride (F K$\alpha$), albite (Na K$\alpha$), anorthite (Al K$\alpha$, Si K$\alpha$, Ca K$\alpha$), microcline (K K$\alpha$), anhydrite (S K$\alpha$), sodalite (Cl K$\alpha$), $TiO_2$ (Ti K$\alpha$), fayalite (Fe K$\alpha$), and SrO (Sr K$\alpha$). All elements were counted for 20 s. Results from these measurements were used in this calibration effort.

In 2022, the 1991 results were cross-checked for validity through 11 new spot measurements performed on another piece of GRR 2409, using a JEOL JXA-8200 EPMA. The experiment was calibrated and performed similar to the 1991 experiment with the following differences. Beam settings of 15 kV and 25 nA and beam defocusing to 10 µm diameter were all used. F-phlogopite (F K$\alpha$) was used instead of fluoride as the F standard. Both F (60 s), K and S (40 s) were counted for durations longer than 20 seconds to enhance count statistics and Na was measured first to limit Na loss.

All analyses utilized the reduction methods described by Armstrong (1982, 1988) and were processed using the CITZAF correction procedure (Armstrong, 1995). Results from these different measurements show remarkable agreement. All elements, except for Fe, F, and S, agree within 10% relative differences (supplementary xx). The difference could be a result of different background fitting methods and sample heterogeneity.

## Measurement Details, Apparatus and Software Processing

X-ray data were collected on the PIXL flight instrument at JPL in a simulated martian environment comprising of martian gas simulant at 6 Torr, -60°C ambient temperature. The Rh anode X-ray tube was operated at flight conditions of 28 kV at 20 µA for all measurements. The sensor head and calibration target box set were positioned as shown in the photo (left) and CAD (right) renderings in Fig. 2. In this configuration the PIXL sensor head was kept fixed while the sample stage moved the calibration targets in X and Y to record the data. OFS imaging and distance measurement of 3 corners of the calibration box were taken to verify that we had consistent instrument standoff across the full measurement range. The X-ray beam boresight was positioned to be orthogonal to the surface of all measured samples.

PTFE and all GRM targets, including the four on the rover calibration target, were measured using 3 × 3 spot grids of 800 s integration per spot, nine spots per detector with grid spot spacing of 0.5 mm. A total real time of 2-hrs per target was commanded. With both detectors recording information, 4-hrs of total integration is recorded across 18 spectra. The small grid pattern was selected to allow any heterogeneity effects on signal response to be both interrogated and subsequently averaged into a single 2-hr equivalent spectrum for each detector. The stepping motor driven stage used to move the calibration target box has an approximate ±2 µm uncertainty and was precise enough to allow commanding of beam placement on each target in all cases.

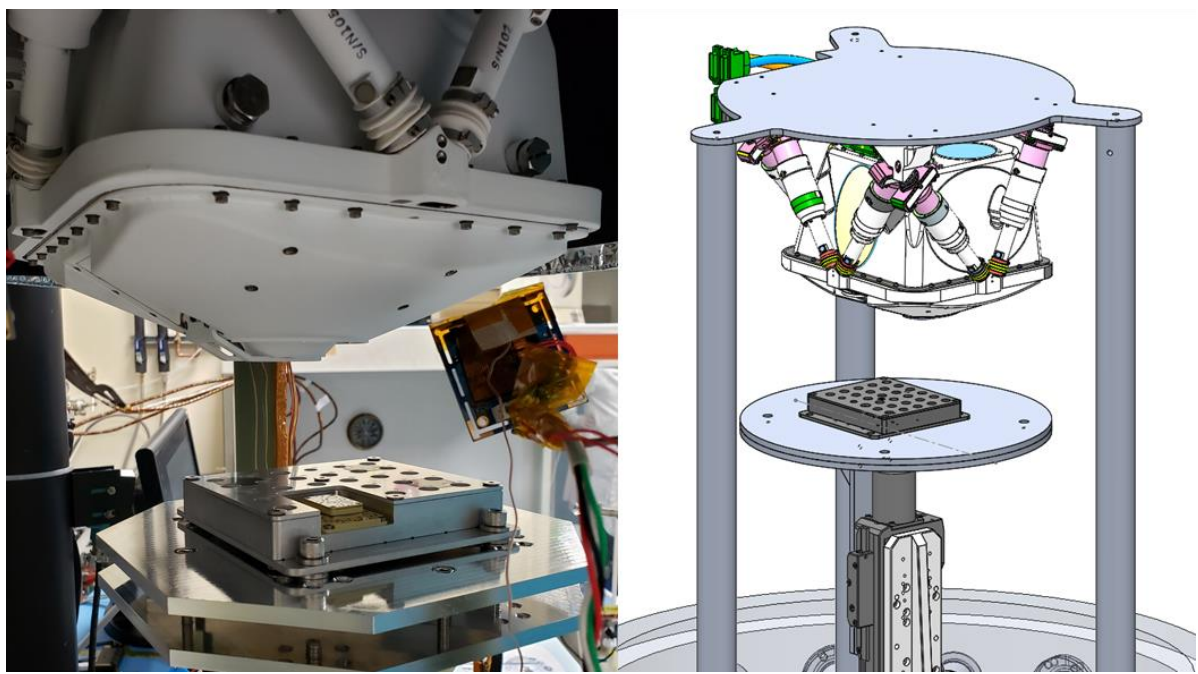

**Figure 2:** Calibration test setup showing a photo (left) of the calibration box in position with the PIXL flight sensor assembly, prior to mounting on the rover arm. The 5-puck rover calibration target is shown in the left photo, inset to the larger calibration box. The rendering (right image) of this test setup shows a broader view of the full chamber setup used.

The pure element and pure compound targets (PTFE excluded,) were measured for 5-min at one integration spot per target. Since these targets were expected to exhibit a high degree of homogeneity and do not require long integration times to resolve trace elements, a single, short duration integration was deemed sufficient.

For all measurements, flight-like data were recorded in May 2019 using in-house software and temperature monitoring of individual components was carried out continuously. Total elemental calibration test time was approximately 36 hours. Following the calibration test, hardware work was deemed necessary to repair faulty parts in the Low-Voltage Control Module (LVCM). The possible effects on the calibration that might have been imparted by this re-work were assessed in January 2020 with additional X-ray measurements made of the rover calibration targets. Due to the nearly identical spectral and count rate responses observed between spectra from the initial calibration test and the re-work test, the original calibration was confirmed to be valid. Nevertheless, the re-work PTFE data was substituted in over the former for use in this pre-flight calibration. Spectra from detector B were used to derive ECFs and the sum of spectra from detectors A and B were utilized for all other targets. For LKSD-4, a noticeable difference in composition of spot no. 7 was observed in the data. Both detector's spectra for this spot were therefore not incorporated along with the remaining 8 spot measurement data in generating a bulk sum spectrum.

Spectra processing from all three steps of the calibration were carried out using PIQUANT v.3.2.6. Raw input and processed output data files involved as part of the full calibration process will be made available on the NASA Geosciences Node Planetary Data System (GN-PDS) (Allwood and Hurowitz, 2021).

## Optic Response Derivation

PIXL's OR was derived using an indirect method that is a modification of our earlier approach (Heirwegh et al., 2018; 2022). This method utilizes the spectral response recorded from PIXL's measurement of PTFE. The method utilizes the PTFE bremsstrahlung background as a means to estimate PIXL's tube plus optic primary emission spectrum; correction to the spectrum to account for primary beam scattering and absorption in PTFE was supplied by PIQUANT. PTFE material was chosen as it exhibits no characteristic X-ray peaks (i.e., it contains no elements with $Z > 10$) of its own and introduces minimal diffraction artifacts in recorded spectra. The PIQUANT software FP approach is also utilized in this method to predict and calculate the full PTFE spectra under varying iterations of the OR profile, including the "unity-optic" case in which photon losses in the optic were assumed to be nil. The OR is realized fundamentally by extracting the difference in bremsstrahlung background profiles derived between measured with-optic PTFE vs calculated no-optic PTFE spectra. This new, iterative approach is described in brief below. A detailed description in this approach is being prepared for publication by Elam et al.. The advantage of the new method over that of Heirwegh et al. (2018) is that it minimizes the impact of the factors (e.g. characteristic lines, low intensity response spectra ends) that diminish the accuracy of Heirwegh et al.'s channel-by-channel approach.

PIXL's OR was approximated using a cubic-spline construct at knot positions of: 0, 4, 6, 8, 10, 12, 14, 16, 18, 20, 25, and 30 keV, assuming an X-ray tube voltage of 28 kV. A PIQUANT FP calculation of a spectrum of PTFE was then made, assuming that the focusing optic response was unity for all transmission energies (i.e., all X-ray photons that entered the optic were focused and transmitted toward the target). The calculated background was then broken up into regions centered on the knots of the spline. The calculated spectrum was fit to the measured spectrum as a set of separate components including the characteristic peaks and the background segments. Fit coefficients were generated at the knot positions. Each coefficient represents the proportional difference between the measured and calculated spectra at each respective knot position. The optic response is dramatically affected in energies near those of characteristic X-ray peaks from the Rh anode of the X-ray tube. For this reason, the peaks were fit separately and their intensities ignored when finding the OR. The fit coefficients were then used to adjust the spline values and a new spline was calculated. This produced a first-iteration guess at a transmission OR. A new PTFE spectrum was then calculated using the first-iteration OR. Like before, a fit of this spectrum to the measured spectrum, ignoring peak amplitudes, permitted calculation of a set of fit coefficients at the knot positions. The new fit coefficients were used to adjust the first-iteration OR thereby refining the approximation of the OR. This process took only 2 – 3 iterations before convergence was realized and the final curve was produced. This approach produces an OR that is not dependent on a precise calculation of the spectrum emitted by the X-ray tube.

## Derivation of Element Calibration Factors

To derive ECFs, each 5 min measurement of the 14 pure element and pure compound targets were processed using PIQUANT's calibrate fitting option. This option operates using the same FP algorithm (eq. 2) employed by PIQUANT to derive elemental concentration for unknown materials. The only difference is that PIQUANT rearranges the FP equation to solve for ECFs as unknowns, treating concentrations of elements in the material as input. Known elemental abundances are therefore used as input by PIQUANT in deriving ECFs.

In practice, one can derive multiple ECFs per element by using multiple reference materials that contain that element. We have found that the most accurate ECFs are derived from measurements of reference materials that have only a few very strong element peaks, such as pure elements or simple compounds. In our earlier work, use of more complex geological reference materials to find the ECFs revealed limitations in the accuracy of calculated ECFs. Many of the elements in complex targets have both lower concentrations and smaller peaks than those found in pure targets. Smaller peaks from multi-element materials often overlap with other peaks or non-gaussian spectral features and are subsequently more strongly affected by errors in the background subtraction. Since knowledge of the element concentrations and degree of target homogeneity is far more certain in pure target materials, this helps to enhance overall confidence in accuracy of derived ECF results.

The only perceived drawback to using pure targets was the occasional presence of intense coherent scatter peaks in spectra due to X-ray diffraction, an effect discussed further in the Discussion Section. The pure target characteristic X-ray lines were in all cases, far more intense than the neighboring diffraction peaks and ultimately, diffraction artifacts did not impact calibration data analysis in any significant way. Therefore, we found that using pure elements or simple compounds to determine the ECFs gave the best results over the entire concentration range. PIQUANT is programmed to interpolate and extrapolate ECFs for other elements not derived for the calibration. Final ECFs are shown in the Results section.

## Assessment of Elemental Accuracy

The accuracy of PIXL in reproducing elemental abundances was assessed primarily by comparing how well the measured abundances compared to the reference compositions. All discrepancies of the calculated versus accepted concentrations were reported as the percentage difference of the measured vs reference values. Accuracy on a per-element basis was then derived using statistical methods, described as follows.

Multi-point spectra from each measured GRM, denoted as $(k)$, were summed to produced one bulk spectrum. Each GRM bulk spectra were processed by PIQUANT's (v.3.2.6) "Evaluate" routine to assess elemental abundance of all PIXL-detectable elements. The PIQUANT calculated results $(calc_{i,k})$ for element $i$ in GRM $k$ were compared to the certificate or reference values $(ref_{i,k})$ as a percentage difference $(PD_{i,k})$, as shown in equation 3.

$$PD_{i,k} = 100 \cdot (calc_{i,k} - ref_{i,k})/ref_{i,k} \qquad (3)$$

Equation 3 represents the discrepancy between the PIQUANT calculated composition and its associated reference (or certificate) value. All $PD_{i,k}$ results were tabulated as a function of their reference concentration and grouped according to concentration $(l)$ ranges, 0 – 0.05%, 0.05 – 0.5%, 0.5 – 5% and 5 – 100%. Data were then further sub-grouped $(m)$ in ranges of element atomic number: $i$ = 1 – 27, 28 – 42, 43 – 56, 57 – 71 and Z ≥ 72. For each such sub-group, we calculated root mean square error of deviation $(RMSD)$; see equation 4,

$$RMSD_{lm} = \sqrt{\Sigma_i \Sigma_k \frac{(PD_{i,k,l,m})^2}{n_{lm}}} \qquad (4)$$

where $n_{lm}$ is the number of data points in subgroup $lm$. The *RMSD* therefore represents the approximate 1-sigma level average deviation expected for all elements found, within an elemental subgroup, for a given range of concentrations. As outlined in the data below, the results from each subgroup are used to place limits at specific mesh points in concentration. Using these mesh points, PIQUANT then interpolates uncertainties for any given concentration of an element that it computes.

The uncertainties derived from this calibration effort therefore represent the bulk of uncertainty associated with any given measurement of any element on Mars. Interpolated RMSD uncertainty values are added in quadrature with the Poisson statistical uncertainty calculated by PIQUANT uniquely for each element $i$ in Mars data. This quadrature sum is then reported as the uncertainty given as output by both PIQUANT, and by extension, PIXLISE programs. More on this is outlined in the Results section.

# Results
## Data Quality
The data recorded in simulated martian environment displayed the behavior expected of a measurement taken in a low-density $CO_2$ atmosphere. All element peaks from Mg and heavier in the periodic table were visible in relevant spectra. Small peaks of Na could also be resolved in most spectra. The XRF spectrum of BHVO2-G recorded across its full data range (0 – 28 keV) is shown in Fig. 5 as an example of the performance delivered by the PIXL instrument. The close up of the 0 – 8 keV region of the same spectrum is found in Fig. 6 to provide a closer view of the major and minor elements peak regions. These data reflect the general quality of data and peak resolution, recorded during this test.

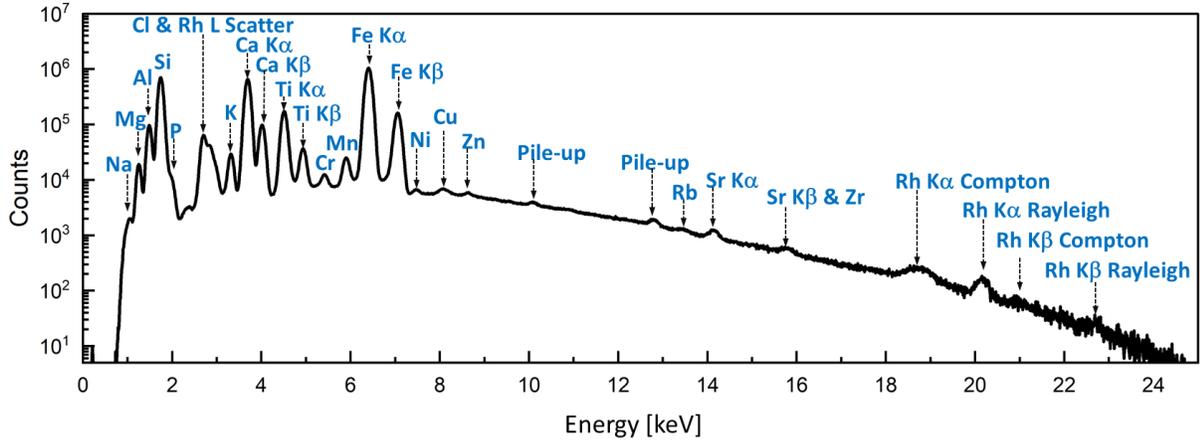

**Figure 3:** Full energy spectrum of BHVO2-G acquired from elemental calibration. Peak width resolution of 154 eV FWHM @ 5.9 keV was observed. Major minor and trace elemental peaks are identified. Counts below 500 eV are due to noise, and are ignored by PIQUANT when quantifying sample composition

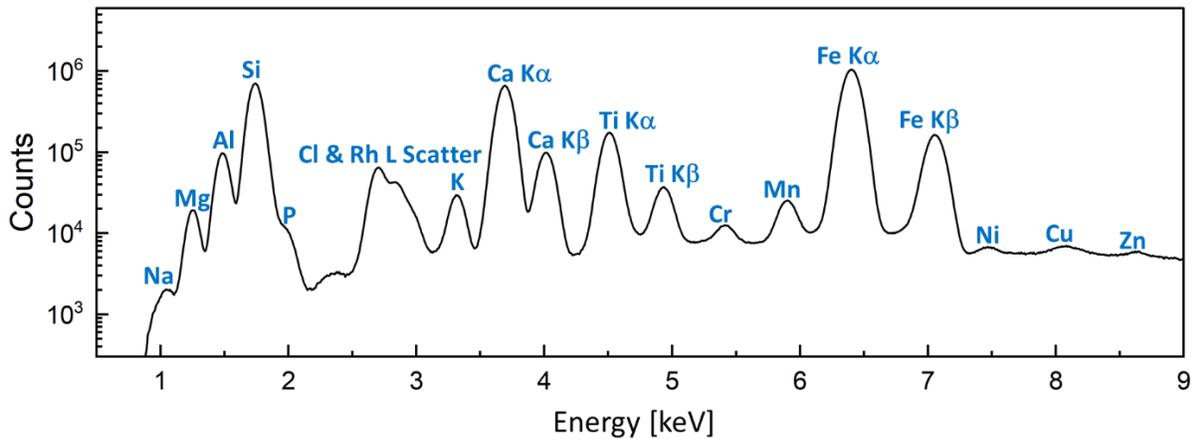

**Figure 4:** Close up major elements portion of the energy spectrum of BHVO2-G acquired from elemental calibration. Major and minor elemental peaks are identified. Peak resolution for this fit is 154 eV FWHM @ 5.9 keV.

## Optic Response

To assess the goodness of the derived OR, described above, a full calibration was run to produce ECFs. Substantial departure of the ECFs from unity was observed for certain elements. The light element ECFs trended downward from unity with decreasing Z from S to Na. We therefore tweaked the OR by raising the cubic-spline 0 keV knot value to be a factor of 2.3 times greater than the value at 4 keV. This brought the lightest element ECFs closer to unity. As well, the knot value at 6 keV was adjusted slightly to bring the Ca ECF closer to unity while also improving the fit to the PIQUANT-calculated background in the vicinity of the Cr K X-ray peaks. The final resulting optic response, now used as part of quantifying unknown materials on Mars, is shown in Fig. 4.

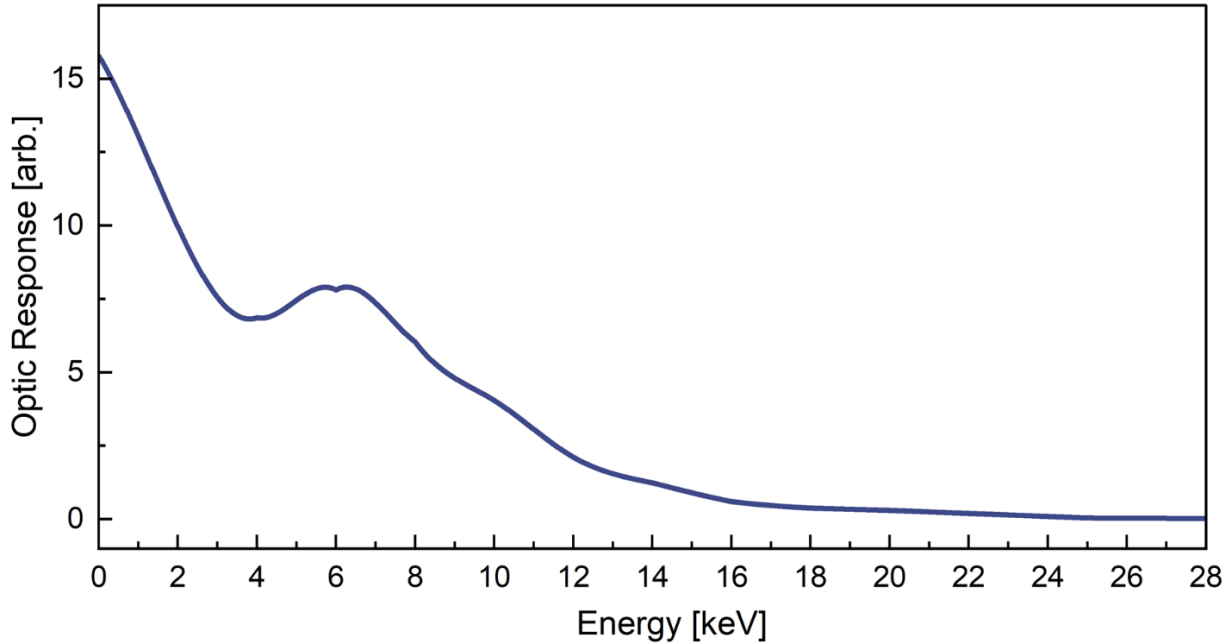

**Figure 5:** Measured unitless transmission function of the PIXL flight OR plotted versus X-ray energy. Described empirical adjustments to the 0 and 6 keV spline points are reflected in this profile.

OR profile values derived here are unitless and the scale was arbitrarily derived as a balance with the value chosen as the solid angle of the optic emitted beam that is incident on the target. This solid angle is denoted as ##INCSR in the PIXL instrument configuration file, (Config_PIXL_FM_SurfaceOps_Rev1_Jul2021.msa). INCSR is adjusted up or down as needed to align the major element ECFs as close to 1 as possible. All ECFs move uniformly up or down by the same factor in this approach. INSCR is ultimately treated as an arbitrary constant in this approach. One might alternatively derive the INCSR, keeping its value unaltered and adjust the OR profile values by the same single constant to align the Ca and Fe ECFs as close to 1 as possible.

## Element Calibration Factors

Using the final tweaked OR, ECFs were rederived and found generally to range from 0.85 to 1.10 about unity (1.0), as shown in Fig. 6. ECFs showing strongest departure from unity were Na and S at 0.75 and 1.14, respectively. The ECF calculated values derived from the 14 pure targets are shown in Fig. 6. Here, dark purple and light green dots represent ECFs derived using only K X-ray and L X-ray peak fitting, respectively. The Zn, Ge, Br, Y and Zr targets all exhibit the emission of both L and K X-ray lines. PIQUANT's ECF generation routine "Calibrate" normally fits both set of lines and generates one ECF only, per element. The resultant ECF reflects the average response from both L and K emissions. Here, we separated the fits of each line set and generated separate ECFs for each to demonstrate the differences that can be observed between different line groupings. Computationally, this is very similar to the unseparated case as PIQUANT averages multiple elemental ECFs into one when quantifying unknowns.

For any element found in a spectrum that does not have a corresponding derived ECF, PIQUANT calculates the missing ECFs using straight-line interpolation between the two nearest known ECFs. The code interpolates using only known ECFs that also appear in the spectrum being quantified. For example, if an ECF for P is being calculated and Al and Ca are the only other elements present in the spectrum, then the code uses only the Al and Ca ECFs to calculate the P ECF. For elements outside the range of known ECFs, the code extrapolates using the same value as the closest known ECF, and again, if and only if that element is also present in the spectrum.

Results for Zn appear at first glance to be slight outliers relative to their measured nearest neighbors. We believe that this is a side effect of attempts to fit pure element spectra without constraining the energy-channel calibration parameters. With so few peaks in some of these spectra, PIQUANT may not always properly converge the appropriate energy-channel calibration and a misfit of the elemental peaks can take place. This would translate into misrepresentation of the true peak area and skew the ECF results. Future efforts will seek to rectify this discrepancy.

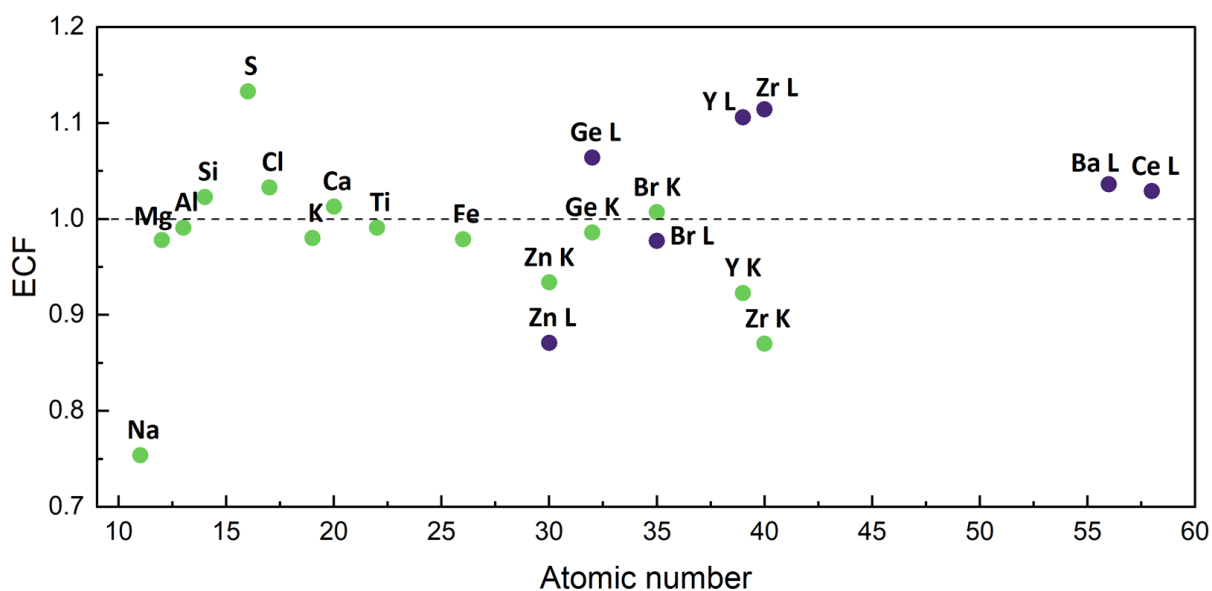

**Figure 6:** Pure element/compound derived ECFs grouped by those calculated using only the K X-ray line emissions (light green) and only the L X-ray lines (dark purple). L X-ray elements: Zn, Ge, Br Y and Zr have corresponding K X-ray peaks appearing in spectra. PIQUANT normally fits both lines and uses the average response to produce one ECF per element. We have however separated Lines for ECF calculations to demonstrate differences that exist in values. PIQUANT recombines multiple entries per element as an average value that is then used in quantification.

## PIXL Elemental Accuracy and Derived Uncertainties

The remainder of the Results section concerns the third step of the calibration. This step serves first as a check of the alignment of the calibration that comes from completing calibration steps 1 (OR) and 2 (ECFs). It then permits derivation of elemental uncertainties on a per-element, per-concentration basis.

Step 3 was assessed using PIXL measurement of all targets, including the GRMs. This third step provides direct assessment of the uncertainties associated with measuring all elements in an unknown material using PIXL. An example of one fit applied to GRM data is shown for JMS-2 in Fig. 7 for the full spectral energy range. A close up of the major and minor element region (0.5 – 8 keV) is shown in Fig. 8. The noise background underlying X-ray peak data is modelled using a combination of fundamental physics calculations in the 0.5 – 7 keV range and SNIP-subtraction in the trace region (7 – 28 keV) (Heirwegh et al., 2022). Gaussian lineshapes are applied to each X-ray line, giving rise to fitted peaks, indicated by different colors in the figures.

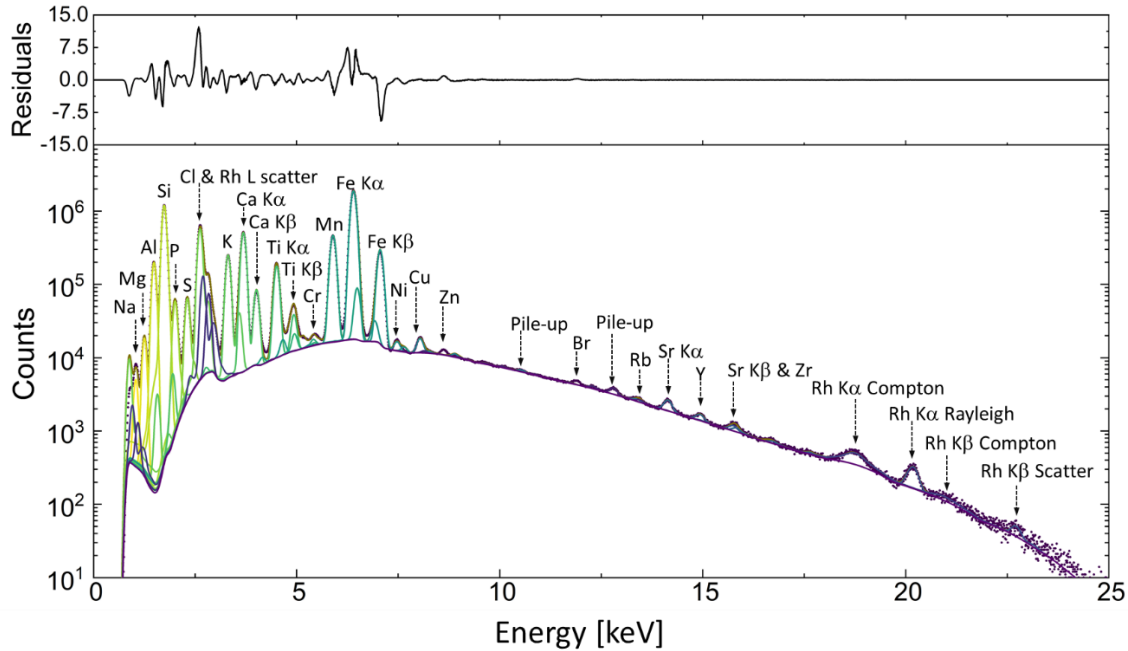

**Figure 7:** Full PIXL XRF spectrum of marine sediment JMS-2 fitted by PIQUANT (v.3.2.6). This spectrum was taken pre-flight by the PIXL flight sensor head hardware. Reduced chi-square fit residuals showing goodness of fit are shown in the top panel.

Percentage difference values were calculated for all visible elements in all calibration target materials. To ensure that only statistically meaningful peak intensities were used, only element abundances found to be greater than 0.01% (100 ppm) were used to evaluate the quantification discrepancies. The resulting discrepancies, grouped as RMSD computations for element atomic ranges and concentrations in GRMS, are shown in Table 3. Results for individual elements are shown in Fig. 9 for the 0.05 to 100% range. We zoomed in on the higher concentration range in Fig. 10 to show the distribution of elements that sit primarily within a ±5% RMSD range. Scatter plot results from Figs. 9 and 10 indicate immediately that the elemental calibration is on the whole, properly aligned; data appear evenly distributed to either side of the line of zero percent different of PIQUANT vs Reference values.

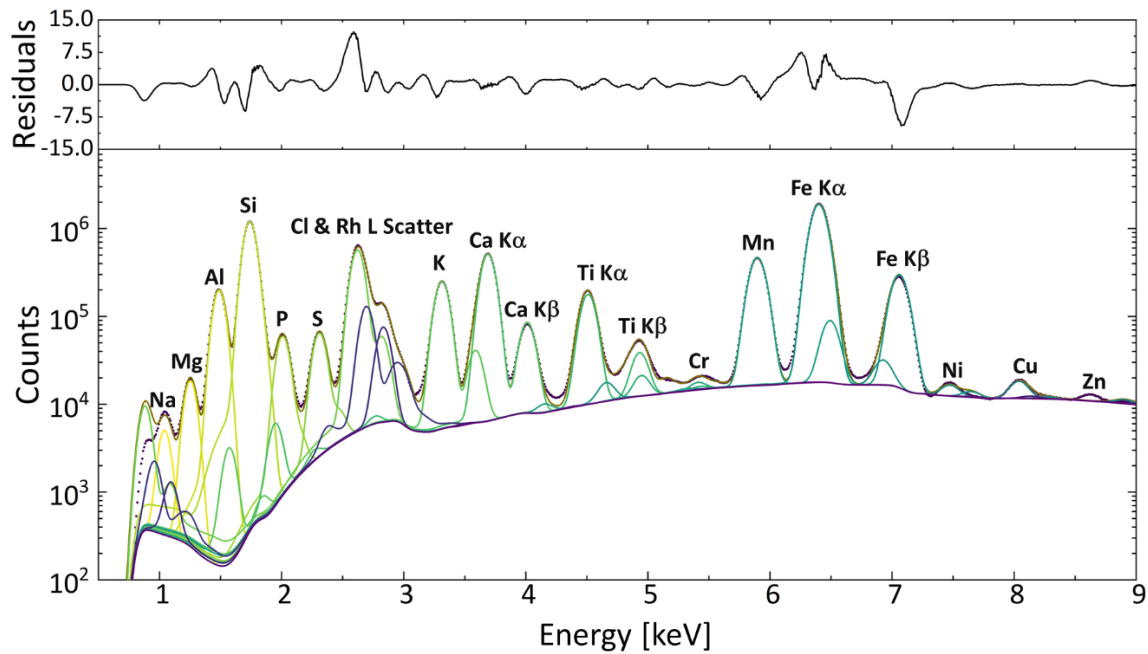

**Figure 8:** PIXL XRF spectrum of marine sediment JMS-2 fitted by PIQUANT (v.3.2.6) in the energy range of most major and minor elements. This spectrum was taken pre-flight by the PIXL flight sensor head hardware. Reduced chi-square fit residuals showing goodness of fit are shown in the top panel.

To assess uncertainties for each element at various concentrations, RMSD groupings were summarized in Table 3. Here, the number of values used *(n)* to calculate each grouping is indicated within. Also shown in Table 3 are the ranges of concentrations used to derive RMSD values as well as the mesh points for the 1 RMSD (1σ) values used to draw the interpolation lines shown in Fig. 9.

In many of the groups in Table 3 there were few to no data points to use to derive a statistically meaningful uncertainty for a given element range at a given concentration. Entries with no data were marked with "--" in the table. We therefore used the RMSD values calculated using the full elemental range of data ($11 \leq Z \leq 92$) within a given concentration group to represent the statistically sparse groups. This application is shown in Table 4, a table that summarizes the final uncertainties expressed for each group.

A few other treatments were applied to several groups in Tables 3, which were annotated and described in the Table 3 and Table 4 captions. The light element ($11 \leq Z \leq 27$), lowest concentration group, indicated as (‡) in Table 3 possessed an RMSD average discrepancy of $3.0 \times 10^3$%. The size of this number reflects the practical limitation on quantifying light major and minor rock-forming element constituents at very low abundances (i.e. < 500 ppm) in a region of the spectrum dominated by intense major element peaks. The use of such a large number as an uncertainty point has little practical meaning. We therefore substituted out this value in Table 4 in place of the RMSD calculated for rest the $Z = 28 – 92$ results in the lowest concentration group.

**Table 3:** Raw RMSD average discrepancies between PIQUANT quantifications and reference values from calibration targets, sorted into groups of concentration (wt.%) and element atomic number. Mesh points used to derive the grid of uncertainties used by PIQUANT and PIXLISE are indicated, as is the concentration ranges used to derive data for each RMSD grouping.

| Mesh Conc. [%] | Conc. Range [%] | RMSD Discrepancies [%] and sample number (n) | | | | | |
|---|---|---|---|---|---|---|---|
| | | Z = 11 – 27, | 28 – 42, | 43 – 56, | 57 – 71, | 72 – 92, | 11 - 92 |
| 5, 100 | >5 - 100 | 4 (45) | 6 (6) | 0 (1) | 0 (1) | -- | 5 (53) |
| 0.5 | >0.5 - 5 | 36 (40) | 1 (1) | -- | -- | -- | 36 (41) |
| 0.05 | >0.05 - 0.5 | 127 (35) | †22 (8) | 232 (4) | †37 (3) | -- | 126 (50) |
| 0 | >0.01 - 0.05 | ‡2952 (25) | 46 (39) | 755 (11) | 87 (15) | 62 (8) | *298 (73) |

† - value not retained due to small sample size. Corresponding Table 4 value derived using all Table 3 percentage difference values across (0.01 - 100%) concentration range for that atomic group. This has effect of increasing uncertainty due to inclusion of lower conc. range elements (>0.01 to 0.05).

‡ - value not used as major element (Z=11-27) quantification for wt.%'s < 0.05 is effectively zero.

* - uses only data from Z = 28 and up; Z = 11-27 data are excluded due to very high errors associated with inability to reliably quantify at this concentration level (expected)

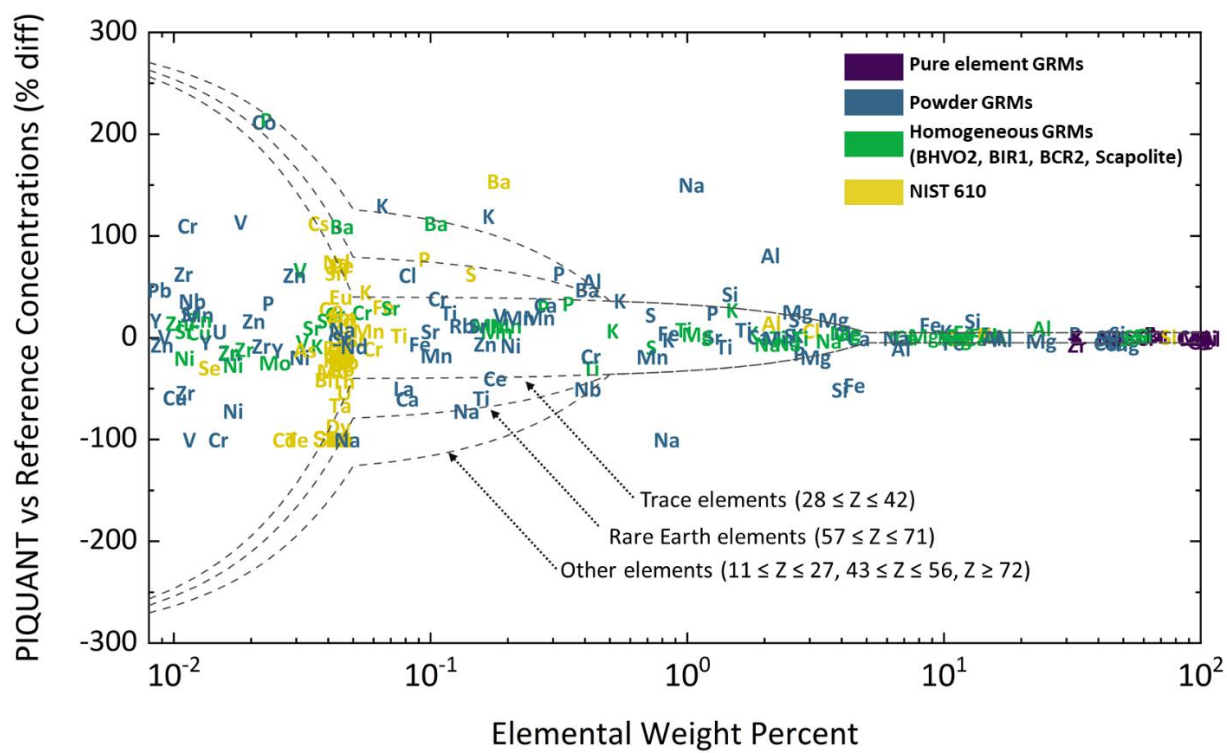

**Figure 9.** Discrepancies between PIQUANT quantifications and reference (certificate) analyses for every element of every GRM plotted as percentage difference vs reference abundance (wt.%). Data are categorized by color according to their origin: being either homogenous, heterogeneous, pure element or compound and the synthetic NIST 610 GRM. Individual discrepancy Elements bound as oxides in standards are plotted as their element-oxide abundance. The grey line segments represent the uncertainties that are returned by PIQUANT, derived from Table 3. Elements are sorted by color in accordance with the nature of their source as either, pure element or compound GRMs, heterogeneous powder GRMs, homogenous GRMs and NIST 610.

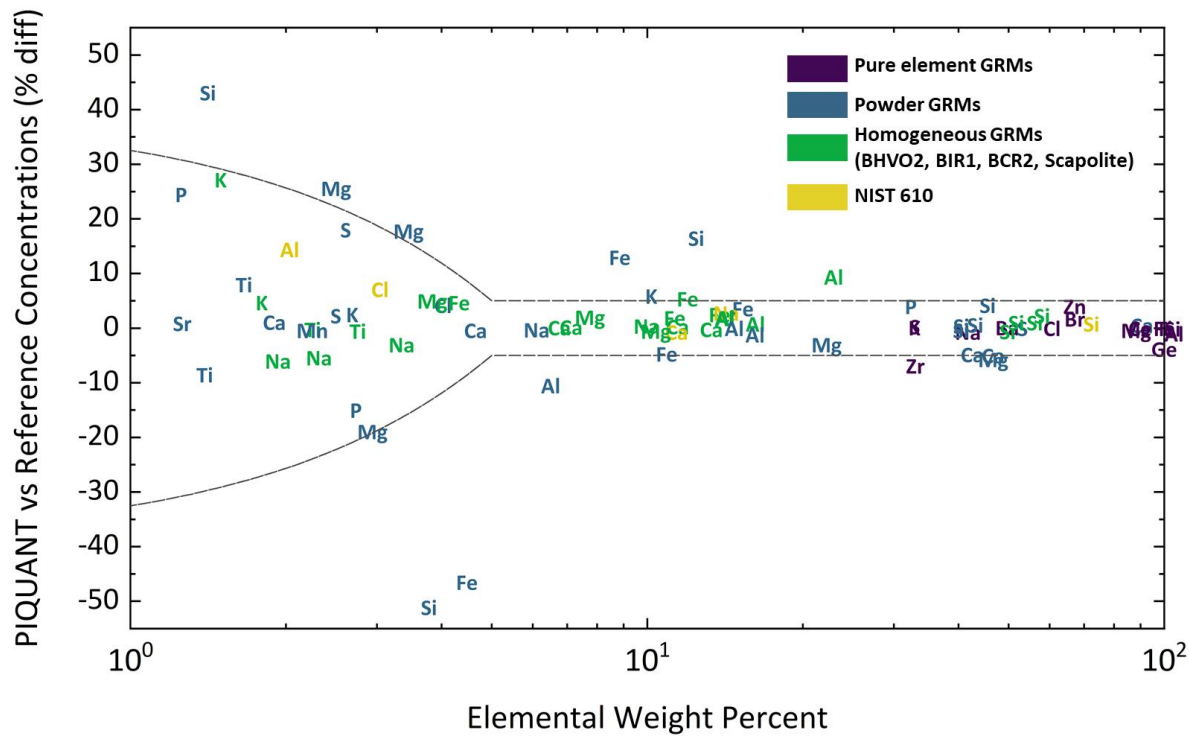

**Figure 10.** Close-up of the 1 – 100% concentration range illustrating spread of percentage difference results about the ±5% RMSD boundaries.

**Table 4:** Final RMSD average discrepancies implemented into PIQUANT and PIXLISE software. Data were derived from calibration target data, sorted into groups of concentration (wt.%) and element atomic number. Mesh points used to derive the grid of uncertainties used by PIQUANT and PIXLISE are indicated as is the concentration ranges used to derive data for each RMSD grouping.

| Mesh point conc. [wt.%] | Conc. Range [wt.%] | RMSD Discrepancies [%] and sample number (n) | | | | |
|---|---|---|---|---|---|---|
| | | Z = 11 – 27, | 28 – 42, | 43 – 56, | 57 – 71, | 72 – 92, |
| 100 | >5 - 100 | 5 *(53)* | 5 *(53)* | 5 *(53)* | 5 *(53)* | 5 *(53)* |
| 5 | >5 - 100 | 5 *(53)* | 5 *(53)* | 5 *(53)* | 5 *(53)* | 5 *(53)* |
| 0.5 | >0.5 - 5 | 36 *(41)* | 36 *(41)* | 36 *(41)* | 36 *(41)* | 36 *(41)* |
| 0.05 | >0.05 - 0.5 | 126 *(50)* | †40 *(54)* | 126 *(50)* | †79 *(19)* | 126 *(50)* |
| 0 | >0.01 - 0.05 | 298 *(73)* | &298 *(73)* | 298 *(73)* | &298 *(73)* | 298 *(73)* |

† - values utilize included full concentration range (0.01 - 100%), has effect of increasing uncertainty relative to Table 3 due to inclusion of lower conc. range elements (>0.01 to 0.05)

& - Applied value of 298% is possibly an overestimate on deviation for this group, given the cleanliness of the bremsstrahlung region in this portion of the spectrum.

The uncertainty values of Table 4 are shown graphically in Fig. 9 as grey dotted lines. As discussed earlier, the final uncertainties reported by PIQUANT for a given element abundance is the quadrature sum of the Poisson statistical uncertainty $(\sqrt{N_i}/N_i)$ with the Table 4 RMSD values. PIQUANT's output uncertainty is calculated according to equation 5.

$$\sigma(PIQUANT_i) = \pm(PIQUANT_i) \times \sqrt{\left(\frac{\sqrt{N_i}}{N_i}\right)^2 + \left(\frac{RMSD_{i,k}[\%]}{100\%}\right)^2} \qquad (5)$$

Here $PIQUANT_i$ is the concentration of element *i* (wt.% in element oxide) calculated by PIQUANT, $N_i$ represents the peak area or intensity of element $i$. The magnitude of the Poisson statistical uncertainty is inversely proportional to the analyzed peak intensity. This is an important contribution to (5) as it permits $\sigma(PIQUANT_i)$ [wt.%] to reflect the statistical noise impact of distinguishing peak from background for any given measurement integration time.

PIQUANT extracts the $RMSD_{i,k}$ value for element $i$ at concentration $k$ by linear interpolation of the built-in mesh from table 4. This component mainly reflects the uncertainty associated with concentration of an element in a sample, for arbitrary integration time. The visual manifestation of this is that uncertainty is interpolated along the grey lines in Fig. 9 as a function of abundance. These uncertainties, reported at 1 RMSD, being based on real measurements of well-characterized standards, represent the best estimate of the uncertainties present in measurements and include all effects encountered in a measurement by the PIXL instrument. We note that although data from less than 100 ppm were excluded entirely, the mesh limits from the lowest concentration elements used were extended to the mesh point of 0%, for completeness. Use of both the Poisson and RMSD uncertainties ensures that uncertainties can be applied to PIXL measurements of any integration time for any elemental abundance.

We note a few other cases of spectroscopic interest. Spectra from the synthetically prepared NIST SRM 610 GRM represented a unique fitting challenge given the many (60+) elements doped at approximate 300 – 500 ppm concentration levels within the glass. In anticipation of potential peak fitting shortcomings that might arise in fitting the dominant L lines of the many rare earth elements, we have shown the results from this target separately (yellow data points) in Fig. 9. The fits for these elements used the L lines, which are smaller and heavily overlapped by the K lines of the major, minor, and trace elements. The quantification of these rare earth elements was expected to be poorer than other trace elements but in fact was exceptionally accurate. The reason for this is probably the large number of L lines for each element that are spread over a fairly wide energy range, making them more independent during the least squares fit.

We further note that non-detections (those having a fit intensity of zero) appeared in the analysis as values of -100%. If these results had concentration abundances that fit within the concentration limits for a RMSD group, then these values were included in all cases. The full calibration file dataset has been made available on the GN-PDS.

Two small errors were observed in this work, post-implementation of the calibration. One was that during generation of the ECFs in the calibration file, the Al$_2$O$_3$ target was converted into a geochemical sum that was greater than 100% (i.e. 100.01%). This error, though now corrected in PIXLISE, was present in the ECF calibration file used to derive the evaluation of materials as unknowns. As a result, PIQUANT was unable to evaluate Al quantifications using the raw Al ECF. Instead, the code utilized nearest neighbor Mg and Si ECFs to calculate an ECF for Al. The interpolated ECF differed from the original Al

ECF, reported in this work, by 1% so was not considered to impact the calibration in a significant way. A second error was that the 15 wt.% water component in the $MgCO_3 \cdot H_2O$ material was denoted as 5% in the standards input file used to derive the ECFs. This discrepancy also was not expected to impact the calibration significantly as the Mg constituent amount was accurately retained at the correct value.

# Discussion

The efforts made to calibrate PIXL to perform accurate elemental analysis have been well-justified. The calibrated PIXL plus PIQUANT analytical system has now been used to analyze numerous PIXL X-ray datasets returned from Mars over the past several years. This FP-based calibration is also one of very few such calibrations ever to be performed on a micro-focused XRF spectroscopy system. This process not only provides the Mars 2020 science team with the means to analyze XRF datasets returned from Mars but also represents a new innovation in advancing quantitative analysis using micro-focused XRF systems. While the present calibration now has heritage use for PIXL, further spectroscopic advances and PIQUANT software upgrades will no doubt be made that will warrant refining the existing calibration. We discuss in this section: a number of factors that impacted the derivation of the existing calibration, factors that are not accounted for by the existing calibration, considerations that an analyst should keep in mind when using the calibration and areas for improvement that might be considered for successive recalibration of the PIXL and PIQUANT as a spectroscopic system.

## PIXL Peak Width Resolution and Influencing Effects

One behavior of PIXL as a spectrometer that must be discussed is the difference in peak width resolution that exists between the ground calibration (148 – 155 eV FWHM @ 5.9 keV) and the post-landing operation (170 – 190 eV FWHM @ 5.9 keV). In general, any significant change in peak resolution has the potential to affect the accuracy in extracting peak areas and by extension, inferring elemental abundances. Large variations (160 – 240 eV) in peak width resolution have been observed in spectra returned by Curiosity Rover's APXS. An investigation into these effects demonstrated correlated trends in analyzed peak areas and abundances as a function of resolution (Pardo, 2015).

For PIXL, peak width was found to increase by about 20 eV following its mounting on the arm of Perseverance and resolution has steadily worsened, albeit on a small scale, since landing on Mars. It was therefore necessary to check the goodness of our reliance on the pre-flight calibration to analyze post-landing datasets. Resolution degradation was attributed primarily to noise introduced by the rover arm cable. Analogue pre-amplifier pulses produced by PIXL's digital signal processor must travel down the 12 m long arm cable, to the rover body analogue front end, to be digitized and sorted. The source of the gradual worsening of the resolution has not been identified but is likely due to continued and ever-present exposure of the SDDs to external radiation sources. Sources include solar and cosmic radiation as well as stray radiation from Perseverance's radioisotope thermoelectric generator (RTG). The goodness and consistency of the ground calibration, as applied to analyzing datasets returned from Mars, is checked periodically about once every 100 sols, primarily through scan and analysis of the onboard BHVO2-G calibration target.

Analysis of PIXL data across a smaller resolution range of 150 – 190 eV has shown (Fig. 2) that concentrations have remained consistent, within statistical variations regardless of the differences in resolution. That we do not see a changing trend in data could be due to our sampling of a smaller range of resolution variation with respect to APXS. Alternatively, the different background subtraction methods employed to analyze each instrument's datasets could also be responsible. APXS data in Pardo (2015) utilized a digital top-hat filter while PIQUANT processes PIXL spectra using the combination of calculated and SNIP background algorithms. Ultimately, as resolution may yet worsen, the PIXL team continues to monitor calibration consistency.

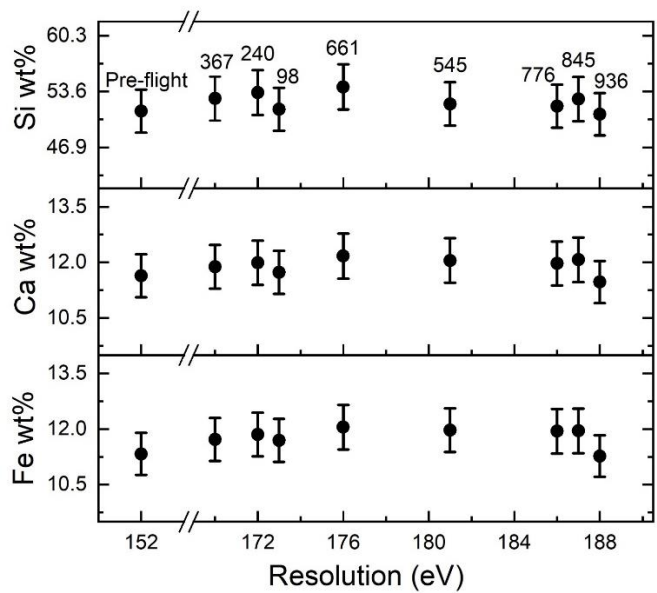

**Figure 12:** Tracking of in-flight 5-puck calibration target quantifications of BHVO-2G Si, Ca and Fe wt.% across varying spectrum resolution. Resolution is assessed at the 5.9 keV FWHM location. The in-flight sol dates in which calibration measurements were recorded are indicated in the top panel.

The difference in peak width resolution and corresponding energy-channel gain necessitated an upgrade of these two default parameter sets in the configuration file used by PIXLISE. This provides generally better convergence of individually fitted single-spot measurements.

## Diffraction Peak Artifacts, Material Heterogeneity and Their Influence in Analysis

PIXL's backscatter geometric design and use of a photon source make possible the detection of photons that are coherently scattered by an interrogated material. PIXL photons incident on mineral grains, having uniformly stacked lattice planes, further give rise to the constructive addition of coherently scattered radiation for specific values of photon energy. The resultant effect is the production of diffraction peaks in PIXL spectra. Peak energies and intensities depend on parameters of the scatter that include the size, orientation and lattice plane spacing of the interrogated crystal. Correlation of diffraction patterns across elemental maps continues be used to infer the mm-scale contiguous nature and size of crystals interrogated by the PIXL (instrument e.g., Tice et al., 2023).

While there is scientific use in analyzing diffraction patterns the overlap of diffraction peaks with characteristic X-rays can hinder accurate elemental analysis. One advantage PIXL has in identifying diffraction peak from characteristic peaks is the fact that it has two detectors. Since diffraction peak position and intensity in spectra is highly determined by the geometry of the incoming and outgoing X-ray beams relative to the crystal orientation, the patterns almost always appear different in spectra as measured between detectors A and B.

Generally, the intensity in spectra of diffraction peaks relative to characteristic lines are the most intense when the signal emerges from a single orientation of a crystal matrix. This is observed in single spot measurements of materials and also in multiple spot measurements made of a larger single crystal (Fig. 12 - top - scapolite). PIXL measurements of large mm- size grains interspersed with other materials, such as that found in the rock target named "Quartier 2" on sol 301, show a reduced relative diffraction intensity relative to the scapolite measurement. However, unique diffraction patterns between detectors A and B are still expressed despite summing multiple spot measurements or more than one mineral phase (Fig. 12 - middle). If on the other hand, one averages multiple measurements of different grains of completely random orientations and further if those grains are much smaller than the beam size, then individual spot measurement diffraction peaks become diminished when these spectra are summed together, as illustrated by the sum spectra taken from the rock target named "Berry Hollow" on sol 505 (Fig. 12 - lower).

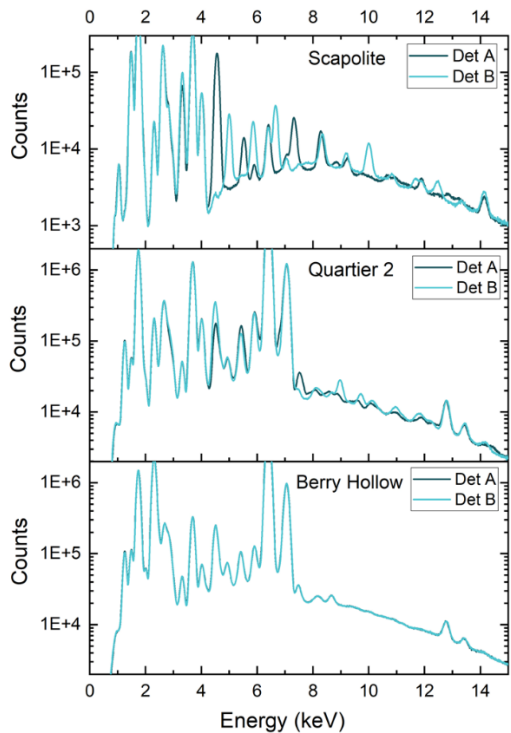

**Figure 12:** Differences in the magnitude of diffraction peak intensities appearing between detectors A and B from bulk summed spectra, depend on the extent that a material produces them and, whether each PIXL-measured spot produces the same pattern. The most severe case is illustrated by scapolite (top panel) as this 2-hr pattern was generated by nine spots all taken of the same contiguous crystal. Here differences between A and B diffraction peak intensities are substantial. A moderate response realized from 17-hr summed measurement of Quartier 2 (middle panel), Crater Floor campaign on Mars includes some mineral mixing but generally similar diffraction patterns summed across multiple spots due to somewhat contiguous crystal nature of rock. Virtually no diffraction is seen in the comparison of A vs B from 17-hr finely grained, amorphously distributed Berry Hollow sedimentary rock patch.

For analysis of spectra using PIXLISE, the challenge of treating and working around diffraction has prompted investigations into both detecting appearances of diffraction in spectra and removing their influence in the analysis pipeline. Presently, PIXLISE employs a machine learning algorithm that was trained to identify all diffraction peaks, and their energy position, appearing across a dataset (Wright et al., 2023) that is based on the difference in peak responses expressed between detectors A and B. A PIXL analyst then has the option to remove these spectra from a region of interest (ROI) that they are analyzing for elemental abundance. An alternative statistical method has also been developed that removes only the diffraction peak from the raw data, thereby avoiding the need to remove whole spectra from summed ROI's. This method utilizes the Wright et al. identification of a diffraction peak in one detector but not the other. The algorithm replaces the diffraction peak data in the first detector using the equivalent channel data, without diffraction, from the second detector's spectrum (Orenstein et al., 2023). While the challenges of analyzing spectra containing diffraction are ever present, these works exemplify the substantial efforts made to mitigate impact of diffraction on quantitative PIXL analysis.

The powder GRMs used in this work demonstrated some small influence from diffraction. Due to the very small crystal size and the averaging of spectra across nine points and 2 detectors, the diffraction peaks observed in individual spot measurements were smoothed into nearly null intensity. No treatment for diffraction was therefore applied in this work and likely any impact to analysis from small diffraction will fold into the derived accuracy limits.

Regarding target material heterogeneity, each of the glass targets provide the analytical advantage of introducing material matrices that are a homogenous mix of all element constituents. The PIQUANT software cannot compensate for multiple material phases within a single quantification run and is only able to assume a homogenous matrix. Also, as their materials are amorphous, the glasses also do not exhibit any evidence for diffraction scatter in the matrix. Glasses are therefore an ideal target set to test for effectiveness of an elemental calibration.

Though the pressed powder GRMs have grains typically much smaller than the interrogation diameter of the beam, visual compositional differences can be observed between spectra recorded at different spots on the pellet. As well, small amounts of surface roughness, even at the level of the roundness of the small grains, can impact accuracy. When multiple GRM measurement spot spectra are averaged, chemical compositions often reflect reference values quite well, though they can have an approximate 2× greater average deviation from reference value relative to the glasses. The powder GRMs should therefore be considered only an approximate representative of a homogeneous matrix. The advantage perhaps in using the powder GRMs for this calibration check is that the uncertainty response is likely more directly applicable to the measurement of the multi-grained, heterogeneous materials found on Mars.

## Validation of the Derived Optic Response and Element Calibration Factors

Derivation of ORs for quantifiable micro-focused XRF systems is a relatively new area of optic research and OR profiles are generally not straightforward to derive. Even with our careful approach, our derived OR and the one produced by Monte Carlo modelling by the XOS supplier show several key differences (Fig. 12); the validities of either curve are not easily assessed. The XOS modelled results show discontinuities at 0.1, 0.5 and 1.8 keV that reflect X-ray absorption in the glass material at the Si L-, O K- and Si K-absorption edges, respectively. In application, these effect below 500 eV are too low in energy to impact our calibration as the beryllium windows on PIXL's X-ray subsystem do not permit transmission of photons at these energies. Our derived OR method does not have the sensitivity in the low-energy region to reveal these effects. Also differing between the two curves is the intensity at the 4 – 6 keV location and other small variations in moving up in energy to 28 keV. Despite their differences however, both OR's possess a similar overall shape across the 0 – 28 keV range.

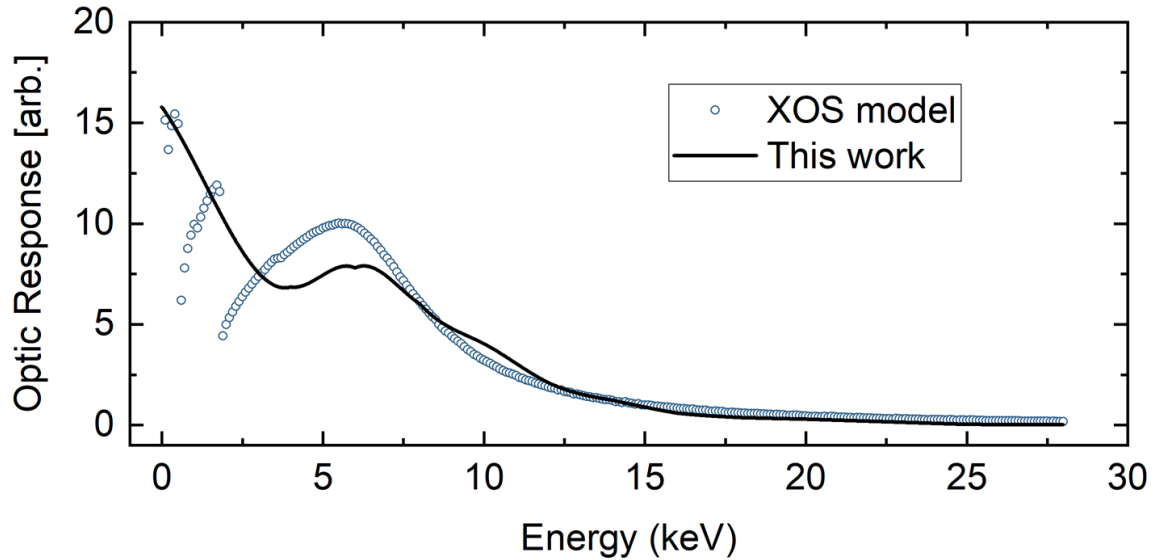

**Figure 12:** Comparison of the flight OR derived in this work to that calculated via Monte Carlo modelling by XOS supplier for the same model configuration. Discontinuous steps in XOS data at 0.1, 0.5 and 1.8 keV reflect photon absorption imposed by the polycapillary glass at the Si L-, O K- and Si K-absorption edges, respectively. ORs were not assessed for transmission intensity relative to each other.

The validity of the OR derived in this work was assessed primarily through the results obtained from the derived ECFs. Strong departure of the ECFs from unity were used to diagnose if issues might be present while minimal departure (~15% or less) was considered acceptable. Two possible divergencies of the OR from ideal were identified and manually corrected for 1), the 0 keV point and 2), the 6 keV point.

Accurate representation of the optic response in the low energy region (0 – 4 keV) is difficult to derive using just the cubic spline method alone. The low energy region background has few counts to utilize in the derivation as most of the photons below 2 keV are absorbed in the beryllium windows of the tube (125 μm) and optic (12.5 μm at both ends). This likely imparts some degree of statistical error in the derivation. As well, many physical and spectral effects can and do impact count intensity in this area, all of which are accounted for in PIQUANT's FP approach. Any shortcomings in PIQUANT's ability to accurately model any of these features would contribute to errors in the cubic spline approach. Features include: the absolute and relative intensities of all lines of the Rh L X-ray complex (2.6 – 3.1 keV) and their associated Rh L escape lines (~1 keV), the detector Compton escape shelf at 0.5 – 1.5 keV (Michel-Hart and Elam, 2017), the bremsstrahlung background predicted by the Ebel X-ray tube model, X-ray tube window thickness and detector geometry and efficiency properties.

Prior to adjusting the 0 keV spline value, the ECFs derived showed a decreasing trend with decreasing atomic number that was sharper than that observed in Fig. 4. Even post-adjustment, a slight decrease of the ECFs with decreasing atomic number is still apparent in the figure. The trend of decreasing instrument coefficients with decreasing Z has also been reported in the calibration of another characteristic X-ray quantification system (Flannigan et al., 2017). In their work, Flannigan et al. included complete FP software corrections for detector geometry and photon absorption. Such corrections would be expected to eliminate any downward trend in instrument constants and yet, the downward trend in their constants persisted. One likely explanation is that correction for all possible sources of photon absorption and efficiency of collection is not always possible. Some detector construction specifications that may impact efficiency are proprietary or, are not fully known and are the subject of ongoing research in detector physics (e.g. Eggert et al., 2006). Efficiency correction issues are likely contributions to the low Z ECF trends seen here.

The trend that the element ECFs rise from Na to S and abrupt drop of the Cl value also merit further discussion. The source of this step or discontinuity between S and Cl may be due to issues with the Ebel (1999) analytical expressions used in PIQUANT to model the characteristic L X-rays emitted by the Rh X-ray tube. These are the X-rays generated from the bombardment of cathode-emitted and accelerated electrons incident on the Rh thick anode in the X-ray tube. Several works have investigated alternative calculation approaches (Finkelshtein and Pavlova, 1999) or improvements (Ebel, 2006) over the Ebel (1999) work. More effort would be needed to identify if the tube model is the cause of ECF discontinuity and further, what should be done to rectify this discontinuity.

The OR derivation presented in this work highlights the challenge of deriving the OR transmission profile as a function of energy. Many effects can impact this derivation. Fortunately, use of ECFs provides almost complete correction of model shortcomings and any inconsistencies in our own derivation and makes reliable quantification possible. One of the achievements of the new OR derivation is that this represents a new, relatively straightforward and innovative advance in calibrating micro-focused XRF systems.

One further note, regarding the ECFs derived in this work, deserves mention. The trend in ECFs shown in Fig. 6 at elements in the range of Fe to Zr and beyond appears to be slightly less smoothly varying across Z than the trend observed in the ECFs derived from the calibration of the PIXL breadboard (Heirwegh et al., 2022), using the exact same target set. The source of this difference was identified following the implementation of the PIXL calibration and determined to be an idiosyncrasy of the spectra fitting algorithm in processing the ECF spectra. When least-squares minimizations are applied to spectra containing very few peaks amidst long stretches of background, the minimization code can struggle to register changes applied to energy-channel and peak width parameters. In some instances, convergence of the algorithm is successfully realized after the algorithm has tweaked the parameters in the wrong direction. This results in slight errors in fitting the elemental peaks and can produce slight errors (<10%) in extracted peak areas. This artifact can be mitigated by choosing a good starting set of calibration parameters, derived from a more complex spectrum (e.g. BHVO2-G) and fixing these for the duration of the ECF fits. This issue was realized and this correction was applied to the fits in Heirwegh et al. 2022), thereby eliminating such variations in the breadboard ECF data.

## Comments on the Accuracy Limits Assessed for PIXL as an X-ray Spectrometer

The calibration presented here is robust and has been checked against measurements made of numerous reference materials. As GRMs were used to define uncertainty limits, the present calibration is optimized for analysis of geological materials containing element abundances typical of rock and mineral materials. This calibration may not be optimal if used on non-geological materials, further illustrated in this work from our need to interpolate and extrapolate to fill in mesh values missing in table 3. As well, use of only a small set of materials meant that individual major elements could not be distinguished reliably to define individual per-element uncertainties. In particular, the Z = 11…27 group all share the same uncertainty value but it is likely that some elements have much lower overall uncertainties than others. Likely, Mg and Si for example, do not share exactly the same uncertainty if assessed across many more materials. Ultimately, more GRMs measured on a PIXL-analogue system might help to fill in some of these less well-defined uncertainty mesh points.

Even more unlikely is that Na would share in this atomic number group uncertainties. Na at highest concentration manifests as a low intensity peak peak at 1.041 keV that is typically dwarfed relative to its nearest neighbor Mg (1.254 keV) and it also overlaps with Cl K-escape and Rh L-escape lines at 0.9 – 1.0 keV and Zn L X-ray lines (1.01 keV). This peak, even with the system resolution realized from this calibration, is difficult to distinguish amidst the background unless there is about 1.5% $Na_2O$ or more in the sample, even with summed spectra that were recorded for time on the order of an hour or more. With existing resolution conditions on Mars being close to 180 eV (FWHM @ 5.9 keV), on average, Na is slightly more challenging to analyze quantitatively. It is therefore questionable if the accuracy limits defined for the more highly expressed elements (e.g. Al, Si, Ca) could be equivalently applied to similar concentrations of Na.

The accuracy limits placed on trace elements ($28 \leq Z \leq 42$) from this work also need discussion. Trace elements of approximately 100 – 500 ppm in this range appear in spectra in areas relatively free of major and even minor element signals. Diffraction interference can also be rather well identified and treated and thus be mitigated from influencing detection and quantification of traces. By crude estimate, most elements Ni – Zr can be detected at the 30 – 60 ppm level and quantified at about 100 ppm or more when diffraction peaks do not overlap a characteristic peak of interest. These limits are applicable to multi-hour summed spectra such as those analyzed in this work or "region of interest" bulk summed spectra from Mars targets. Minimum limits increase as integration time decreases. New work to model PIXL spectra has been conducted that demonstrate limits applicable to our most typical single-spot integration time of 10 s, as might be found in the average Mars basaltic target (Christian et al., 2023). A reader may reference this work to gain an idea of the limits that may be relevant when available X-ray statistics are at a minimum.

We identify that the accuracy limits on longer duration spectra, placed by this calibration on certain trace element RMSD groups are highly conservative and may overestimate the true uncertainties. This is a potential consequence of using extended ranges to infer these trends. Possible RMSDs affected include the Table 3 atomic number groups ($28 \leq Z \leq 42$) and ($57 \leq Z \leq 71$) in each of the 0.01 – 0.05 and 0.05 – 0.5 wt.% ranges. The original $28 \leq Z \leq 42$ RMSD results of 22% (0.05 – 0.5 wt.%) and 46% (0.01 – 0.05 wt.%) likely reflects the true uncertainty much more strongly than the averaged values of 126% (0.05 – 0.5 wt.%) and 298% (0.01 – 0.05 wt.%), shown in Table 4. The more conservative estimates in Table 3, indicated with a dagger (†) symbol were derived using an average of the entire concentration range of values (0.01 – 100%) for the entire group. By folding in the estimates from the lowest concentration groups (0.01 – 0.05%) we artificially inflate the estimates at (0.05 – 0.5%). If this is true, then the uncertainties applied to many trace elements are likely overestimated. A similar observation exists for the $57 \leq Z \leq 71$ RMSD groups. A possible future fix to this would be to use the exact limits found in Table 3, for these four affected groups. Increasing the GRM sample size might help to improve statistical analysis here as well.

We mentioned earlier the many possible contributions of uncertainty that go along with performing PIXL measurements on materials. This calibration folds in most of these with a few exceptions. Exceptions include surface roughness and instrument tilt effects. Neither were factored into the existing calibration and in practice, these effects have been observed in PIXL measurements to introduce highly significant effects on X-ray peak intensity and detector response differences in measured data. Alternative treatment and post-processing mitigation strategies have been developed by the PIXL and PIXLISE team to prevent spectra exhibiting these behaviors from impacting analysis. Therefore, the existing calibration is generally considered to be valid in the treatment of the unaffected spectra.

One other possible impact to the uncertainties and goodness of the calibration was considered in this work. In 2018, an international congress voted to redefine the fundamental SI unit definitions to be derived entirely by known physical constants rather than relying on physical objects and materials that are subject to change over time (NIST, 2019). This rederivation resulted in several units changing in value by no more than about 0.002% relative to the pre-2018 tabulations. These differences were small enough that their impact to this calibration, the uncertainty limits or the effects on the derived FP databases would be insignificant if the new values were to have been updated as part of this work.

## Statistical Considerations and Outliers

The RMSD uncertainties presented above were produced using statistical analysis assuming a normal distribution. All uncertainties in the tables reflect the 1-sigma spread in observed data. This means that we are confident that 67% of all PIXL-derived elemental abundances from unknown materials will fall within these 1-sigma (1-RMSD) limits. If we use $SiO_2$ in the 5 – 100% range as an example, the 1-RMSD value is ±5%. Extending to 2-RMSD (2σ), 95% of abundances would fall within double the 1-RMSD limits (i.e. ±10%) and likewise 98% of data would fall within three times these limits (i.e. 3-RMSD at ±15%). The key message here to note is that accuracy limits are not absolute but are rather based on statistical probability. Strong departures of agreement of PIXL+PIQUANT with respect to true composition are most likely introduced by effects such as material heterogeneities, or multi-phase beam straddling and surface roughness effects. From the present calibration departures may also be due, in part, to inaccuracies in the reference compositions used.

One point under consideration for future iterations of PIXL's possible re-calibration is that the 14 pure targets, used to derive ECFs, were in turn used in step three, being treated as unknowns. This introduces an inherent circular bias that may influence or shrink somewhat the accuracy limits derived in this work. Future efforts will introduce a limitation on the use of the 14 targets to prevent any amount of statistical bias.

We point out several such "outlier" discrepant data points that arose from our analysis (Table 5). In this table, we identify results for elements possessing compositions of significant levels (>0.5 wt.%) in which deviations from quoted reference compositions were greater than 20%. We note that these values were included as part of the RMSD calculations described above.

**Table 5:** Outliers calculated for targets possessing elemental concentrations of 0.5 wt.% or greater and discrepancies exceeding 20% relative differences. All discrepancies occur in the powdered GRMs or scapolite, primarily when concentrations of major elements are unusually small relative to typical GRMs

| Element | Standard | Ref. wt. % | Discrepancy (rel. %) |
|---------|----------|------------|----------------------|
| Na | LKSD-4 | 0.7384 | -100 (N.D.) |
| Na | SRM 694 | 0.9205 | 150.4 |
| Mg | GYP-B | 2.3339 | 25.6 |
| Al | SRM 694 | 1.9267 | 81 |
| Si | COQ-1 | 3.6427 | -51.1 |
| Si | GYP B | 1.3615 | 43.1 |
| P | JMS-2 | 1.2529 | 24.5 |
| S | JMS-2 | 0.7201 | 22.7 |
| K | scapolite | 1.4937 | 27.2 |
| K | SRM 694 | 0.5459 | 36.5 |
| Fe | COQ-1 | 4.4782 | -46.5 |

N.D. – not detected, element was assessed for quantification but no peak could be found.

We further note that all of these errors occur in the 7 powder standards except for K in scapolite (which has several large diffraction peaks in each spectrum). These outliers are due presumably to inhomogeneities of some kind in the standard specimen or to the presence of diffraction peaks in the spectrum. The Na outlier points may also be due to poor background removal in the lowest part of the spectrum near the low energy cutoff of the detector signal processing.

It is also possible that the discrepant results for Mg, Al and Si have occurred due to the unusual nature of fitting spectra with both unusually low concentrations of these elements and having peaks in close proximity to intense neighboring lines from S (Gyp-B) or P (SRM 694). PIQUANT includes functions to represent peak "tails" that arise due to incomplete charge collection effects (Scholze and Procop,, 2009) though these functions have not yet been rigorously assessed for accuracy. If any low concentration outlier elements are superimposed on these tails and the tail functions are not fully accurate, then errors in extracting element peak areas are likely to occur.

Another possibility may be as follows. As it is challenging to fit low concentration light element peaks, it may be that better quantitative results might be realized with a separate set of ECFs catered to analyzing low concentration elements. We consider that ECFs were derived using targets possessing very high concentrations of the elements of interest. It is not necessarily the case that the same ECFs are ideal for use on very low concentrations of the same elements, in certain cases. More work would be needed to evaluate these situations.

### X-Ray Data Analysis Considerations

We discuss and summarize finally a few points for consideration that should be made when analyzing PIXL X-ray data using the PIQUANT calibration. First, statistically, individual per-spot measurements taken from rocks on Mars have inherently high uncertainties for most elements due to the short integration time (~10 s). Summing multiple spectra together first and then analyzing the single bulk summed spectrum can enhance the signal to noise ratio and reduce the contribution from the Poisson statistical uncertainty. Greater confidence may then be placed in the returned abundance results. This is an encouraged practice when quantifying regions of interest that are found to bear a single mineral type.

Second, some elements are difficult either to fluoresce (e.g. Na) or to detect and quantify (e.g. Ce), specifically if they are in regions of the spectrum in which significant overlap with other characteristic X-ray peaks or diffraction peaks occur. The concern is that in cases such as these, PIQUANT may return a positive detection for an element that is not actually there. The element Ce is one example in which and quantifications at the 100 – 500 ppm level, derived from measurements of targets found on Mars, need yet to be verified (Christian et al., 2023). Ce is measured using its L X-ray lines (4.8 – 6.1 keV) which are superimposed on the dominant K X-ray emissions by common rock forming elements of Ti (5.4 keV), Cr (5.4 keV), Mn (5.9 keV). All of these peaks further sit atop a calculated background that may have small but significant errors that could impact accurate extraction of Ce peak area. Database descriptions of L X-ray lines are also prone to greater levels of error in their intensities and peak positions, thereby further reducing confidence in positive detection. More effort is needed to enhance our confidence in validating positive detections of low levels of Ce.

Last, PIQUANT only quantifies those elements that were selected for assessment by an analyst in the first place. An element may be present in a given measurement but unless that element was selected for as part of the PIQUANT quantification, it will not be found or indicated in any way in the PIQUANT output. A false negative detection might be interpreted in this case. Practically, PIQUANT processing time increases proportionally as the square of the number of elements in the sample set. It is therefore computationally expensive to include many more elements than are needed to interrogate the sample. A recommended practice is to include a rather standard limited element set and to only add additional elements if they are geologically relevant or if spectra indicate that others might be present.

# Concluding Remarks

PIXL comes from a long line of calibrated XRF spectrometers sent to Mars and, new to this instrument, brought with it an immensely powerful capability of quantitative elemental mapping. Preparing the completed PIXL hardware in effort to meet its scientific potential required rigorous steps toward its calibration. This required the building of software tools, formation of a calibration protocol, testing made using carefully selected materials set, full evaluation in a simulated martian environment test setup and dissemination of the calibration results.

PIXL was calibrated using a three-step physics FP approach that included first, derivation of the X-ray optic response and second, calibration alignment (ECFs) using a few pure materials. Quantification alignment uncertainties were assessed in the third step using the pure materials and additional GRMs. Both the alignment ECFs and uncertainties are a part of the full calibration package used by PIQUANT and PIXLISE. This calibration has been implemented in the analysis of PIXL's returned X-ray data since Perseverance's landing on Mars. The calibration uncertainties can be applied to any measured quantity of element abundance recorded across any duration of integrated measurement. The total PIQUANT per-element uncertainty is the quadrature sum of the RMSD results arising from target set measurements with the Poisson peak statistical uncertainty. When Poisson uncertainty is very small, the sum uncertainty reflects the 1-σ spread in the target set measurements, for a given elemental concentration.

This work further illuminated a few characteristics of the data analysis process that could externally impact accurate and direct use of the existing calibration. Characteristics include: limitations in our understanding of the X-ray tube model, errors introduced by rock surface roughness and instrument tilt, uncorrected diffraction peaks, RMSD data paucity for certain GRM concentration and atomic number ranges, and limitations in our ability to place confidence in the reality of positive detections returned by PIQUANT for certain elements.

The magnitude of the significance of this calibration effort is three-fold. First, this calibration makes PIXL's remote geochemical analysis of Mars possible. Second, the efforts to calibrate PIXL have led to new innovations and technique developments that will support the growing research field of micro-focused XRF quantitative analysis. Third, this work identifies several small weaknesses of the calibration that might be addressed by future efforts.

## Acknowledgments
WTE and BCC gratefully acknowledge the impact of W. C. Kelliher for his vision of the future of XRF on Mars and facilitating the collaboration which led to our involvement with the development of PIXL. The authors extend their appreciation to: Professor Penny King, ANU, for providing the powder GRMs used in this work, Dr. Clay Davis, NIST, for donation of the NIST 610 material, Dr. Steven A. Wilson, formerly of USGS, for valuable assistance in interpreting GRM certificate compositional data and, Ning Gao, XOS, and XOS for supplying Monte Carlo modelled optic response data. The authors also thank Professors Allan Treiman, LPI, and David A Klevang, DTU, for manuscript editing support.

## Funding
This work was funded and carried out under contract with the National Aeronautics and Space Administration [80NM0018D0004].

## Statements and Declarations
The authors have no relevant financial or non-financial interests to disclose. The authors have no competing interests to declare that are relevant to the content of this article. All authors certify that they have no affiliations with or involvement in any organization or entity with any financial interest or non-financial interest in the subject matter or materials discussed in this manuscript. The authors have no financial or proprietary interests in any material discussed in this article.

# CRediT

Conceptualization: WTE, CMH, LGA, NB, BCC III, JAH, ACA, LAW; Data curation: MES, RWD, PRL, RR, JHK, MHA, AK, RAR, CH, MSA, LAW; Formal Analysis: WTE, CMH, MCF; Funding Acquisition: ACA, LAW, MCF, RAR; Investigation: CMH, WTE, YL, LAW, ACA, JAH; Methodology: CMH, WTE, LGA, NB BJN, CH; Project Administration: LAW, MCF, RAR, ACA, JAH; Resources: YL, ACA, GRR, MSA; Software: WTE, CMH, LPO, KPS, ACA; Supervision: ACA, MCF, LAW, JAH, RAR; Validation: CMH, WTE, MCF; Visualization: AD, CMH, CH, WTE; Writing – original draft: CMH, WTE, AD; Writing – review & editing: WTE, CMH, LAW, YL, JAH, GRR, MCF, CH